\documentclass{aa}  

\usepackage{graphicx}
\usepackage{txfonts}
\usepackage{upgreek}
\usepackage{gensymb}
\usepackage{chemformula}
\usepackage{enumitem}
\usepackage{multirow}
\usepackage{ragged2e}
\usepackage[switch, modulo]{lineno}

\expandafter\def\expandafter\normalsize\expandafter{%
    \normalsize%
   \setlength\abovedisplayskip{3pt}%
    \setlength\belowdisplayskip{3pt}%
    \setlength\abovedisplayshortskip{3pt}%
    \setlength\belowdisplayshortskip{3pt}%
}

\bibpunct[:]{(}{)}{;}{a}{,}{,}

\usepackage{hyperref}

\begin{document}

   \title{A fast machine learning tool to predict the composition of interstellar ices from infrared absorption spectra}

   \author{Andr\'es Meg\'ias \and \inst{1,2}
          Izaskun Jim\'enez-Serra \and \inst{1}
          François Dulieu \and \inst{3} 
          Julie Vitorino \and \inst{3}
          Bel\'en Mat\'e \and \inst{4} \\
          David Ciudad \and \inst{5}
          Will R. M. Rocha \and \inst{6}
          Marcos Mart\'inez Jim\'enez \inst{1} \and
          Jacobo Aguirre \inst{1}
          }

   \institute{Centro de Astrobiolog\'ia (CAB), CSIC-INTA,       Carretera de Ajalvir, km 4, 28850, Torrej\'on de Ardoz, Spain\\
              \email{$\,$amegias$\,$@$\,$cab.inta-csic.es}
         \and
            Facultad de Ciencias F\'{i}sicas, Universidad Complutense de Madrid, 28040, Madrid, Spain
         \and
             LIRA, CY Cergy Paris Universit\'e, Mail Gay-Lussac, 5, 95000, Neuville-sur-Oise, France
        \and
            Instituto de Estructura de la Materia (IEM), CSIC, Calle Serrano, 121, 28006, Madrid, Spain
        \and
            MasOrange, Paseo del Club Deportivo, 1, 28223, Pozuelo de Alarcón, Spain
        \and   
            Laboratory for Astrophysics, Leiden Observatory, Leiden University, PO Box 9513, NL 2300 RA, Leiden, the Netherlands
             }

   \date{Received December 31, 2024; accepted August 14, 2025.}

 
  \abstract
   {Current observations taken by James Webb Space Telescope (JWST) allow us to observe the absorption features of icy  mantles that cover interstellar dust grains, which are mainly composed of \ch{H2O}, \ch{CO}, and \ch{CO2}, along with other minor species. Thanks to its  sensitivity and spectral resolution, JWST has the potential to observe ice features towards hundreds of sources at different stages along the process of star formation. However, identifying the spectral features of the different species and quantifying the ice composition is not trivial and requires complex spectroscopic analysis.}
   {We reduce the difficulty and the time employed in doing this task by developing a new software analysis tool based on machine learning.}
   {We present Automatic Ice Composition Estimator (AICE), a new tool based on artificial neural networks. Based on the infrared (IR) ice absorption spectrum between 2.5 and 10 $\upmu$m, AICE predicts the ice fractional composition in terms of \ch{H2O}, \ch{CO}, \ch{CO2}, \ch{CH3OH}, \ch{NH3}, and \ch{CH4}. To train the model, we used hundreds of laboratory experiments of ice mixtures from different databases, which were reprocessed with baseline subtraction and normalisation.}
   {Once trained, AICE takes less than one second on a conventional  computer to predict the ice composition associated with the observed IR absorption spectrum, with typical errors of $\sim$3 \% in the species fraction. We tested its performance  on two spectra reported towards the NIR38 and J110621 background stars observed within the JWST Ice Age program, demonstrating a good agreement with previous estimations of the ice composition.}
   {The fast and accurate performance of AICE enables the systematic analysis of hundreds of different ice spectra with a modest time investment. In addition, this model can be enhanced and re-trained with more laboratory data, improving the precision of the predictions and expanding the list of predicted species.}


   \maketitle
%

\section{Introduction}

Interstellar dust particles are known to be constituted of a silicate-rich core, which may also contain carbonaceous material \citep{Draine2004, Draine2007}. With the cold temperatures reached in the interstellar medium (ISM), being as low as $T \simeq 10\;{\rm K}$ in molecular clouds and starless cores, volatile compounds present in the gas phase can condensate onto dust grains and react on its surface, forming so-called icy mantles \citep{Boogert2015}. This process gets accelerated during the prestellar stage, when the core collapses and its central gas densities reach high values $\gtrsim\,$$10^5$--$10^6$\ cm$^{-3}$ (e.g. \citealp{Caselli2002}). Apart from the depletion itself, which provides carbon monoxide (CO) and atomic species such as hydrogen (H), carbon (C), oxygen (O), and nitrogen (N) onto grains, surface chemistry plays an important role, as it drives the formation of other species such as water (\ch{H2O}), carbon dioxide (\ch{CO2}), methane (\ch{CH4}), and methanol (\ch{CH3OH}). These species form thanks to the high mobility of H atoms, which hydrogenate simple species such as O, C, N, or CO, yielding \ch{H2O}, \ch{CH4}, \ch{NH3}, and \ch{CH3OH}, respectively \citep{Dulieu2010, Quasim2020, Redaelli2023, Watanabe2002}. Further energetic and thermal processes of the ice may occur along the process of star formation (\citealp{vanDishoeck2006}; \citealp{Boogert2015}; \citealp{Herbst2022}; and references in them), inducing a grain chemical structure that may contain interstellar complex organic molecules (COMs), which can be transferred to planetesimals and cometesimals \citep{Altwegg2016}.

However, the characterisation of the chemical composition of ices is challenging, since it requires observations of ice absorption features in the infrared (IR) domain against bright enough background sources. These signatures are due to the bonds and functional groups present in the molecule \citep{Boogert2015, Rocha2022}, which cause wide spectral features that are often shared by different species, further complicating the identification of complex solid mixtures due to degenerate solutions. From ground-based observatories, the main components of the icy mantles have been measured towards different dense molecular clouds and dense cores: water (\ch{H2O}), carbon monoxide (CO), carbon dioxide (\ch{CO2}), and methanol (\ch{CH3OH}), as reported in the literature (e.g. \citealp{Guillett1973, Whittet1989, Boogert2013, Goto2021}). Other minor species with fainter absorption features have also been detected, such as the \ch{^{13}C} isotopologue of carbon monoxide (\ch{^{13}CO}; \citealp{Boogert2002}), carbonyl sulfide (OCS; \citealp{Geballe1985, Palumbo1997}), and the cyanate ion (\ch{OCN-}; \citealp{Lacy1984, Pontoppidan2003, vanBroekhuizen2005}). However, achieving successful observations from the ground is not straightforward due to the presence of telluric lines from Earth's atmosphere and to high noise levels in the IR. In this regard, space telescopes such as ISO (Infrared Space Observatory), AKARI, and Spitzer have enabled higher-quality ice absorption observations (e.g. \citealp{Schutte1996, Boogert2011, Noble2013}), which yielded the detection of solid methane (\ch{CH4}), ammonia (\ch{NH3}), and the $^{13}$C isotopologue of carbon dioxide (\ch{^{13}CO2}). Additional features have been tentatively associated with molecules such as \ch{H2CO} or HCOOH \citep{Schutte1999, Keane2001}. The number of sources successfully observed with these telescopes remains low due to the limited sensitivity of these instruments. In addition, the observed spectra either lack a wide-enough wavelength coverage or the spectral resolution was scarce, limiting the study of the ice chemical composition from a systematic and comprehensive point of view thus far.

Thanks to its high sensitivity and unprecedented spectral resolution, the James Webb space telescope (JWST) is providing the most detailed ice absorption spectra ever observed. \cite{Yang2022} reported the first view of the ice composition as seen by the JWST towards a young Class 0 protostar, finding features of \ch{H2O}, \ch{CO2}, \ch{CH3OH}, \ch{CH4}, and \ch{NH3} (the CO band at 4.67 $\upmu$m was not covered). \cite{McClure2023} and \cite{Dartois2024} reported the IR absorption spectra towards the two highly extinguished background stars NIR38 and J110621 in the Chamaeleon I dense molecular cloud, as part of the Early Release Science (ERS) program Ice Age. Besides the major species mentioned above, these authors have also found traces of minor compounds such as \ch{OCN-} and \ch{OCS} as well as the \ch{^{13}C} isotopologues of CO and \ch{CO2} \citep{McClure2023, Brunken2024}. The dangling bonds of OH have also been identified thanks to JWST IR absorption spectra \citep{Noble2024},  offering details of  the physical structure of the ice. \cite{Rocha2024} observed two young protostars with JWST, obtaining detailed spectra with several absorption features, some of which have been associated with molecular species and anions such as \ch{HCOO-}, \ch{CH3CHO}, and \ch{CH3OCHO}. In all the studies above, the identification of the different species is the result of a detailed and complex spectroscopic analysis, which in most cases had to be carried out band-by-band. In addition, this identification is done by eye and the observed molecular bands are integrated and divided by the experimentally derived band strengths of pure ices in order to obtain the ice column density (see e.g. \citealp{Bouilloud2015}). However, it is well-known that the shape (and even the position and intensity) of the IR molecular bands is affected by the ice mixture and ice temperature, which adds significant uncertainties to the accurate quantification of the ice chemical composition \citep{Boogert2015, Rocha2022}.  Although there are measurements of molecular band strengths in ice mixtures, it is difficult to properly characterise the exact proportions for the possible mixtures containing a certain species. That is why it is common to use pure ice spectra as a first approximation.

With the aim of improving and facilitating the determination of the composition of ices, \cite{Rocha2021} presented a machine learning tool called \textsc{Eniigma} which uses genetic algorithms to find the combination of experimental ice spectra that best fits the observations. This software has successfully been used for analysing JWST data in the works of, for instance, \cite{McClure2023} and \cite{Rocha2024}, providing a detailed analysis of the possible ice components and the calculation of ice column densities using experimental band strengths. Genetic algorithms, however, are computationally expensive, requiring long computational times to search for all possible ice mixture combinations of laboratory spectra. Given that JWST promises the delivery of ice absorption spectra towards hundreds (if not thousands) background sources (see e.g. the case of the Chamaeleon I cloud in the Ice Age program; \citealp{McClure2023}; \citealp{Smith2025}), faster and automated alternative approaches need to be developed to efficiently analyse the ice chemical composition of large source samples.

In this work we present a new and publicly available machine learning tool, Automatic Ice Composition Estimator (AICE), which is based on artificial neural networks. AICE is a fast and easy-to-use tool programmed in Python language, which has been trained using ice laboratory data. Once trained, AICE enables the analysis of hundreds of ice absorption spectra in a matter of minutes (one spectrum is analysed in less than one second). Therefore, AICE provides the opportunity to carry out large and complex IR ice absorption spectra in an automatic and systematic way, enabling the study of ices using statistically significant samples.

The paper is organised as follows. In Section 2, we present our model in detail, describing the architecture and dataset used. In Section 3, we present the results of the training of the model and of the test with real JWST data that have already been published. In Section 4, we discuss the different sources of uncertainty present in our tool and compare it with other methods. Finally, in Section 5 we summarise our work and present our conclusions with possible future steps.

\section{Methods: Design and training of the AICE tool}

AICE is a Python 3 analysis tool that consists of several artificial neural networks trained on laboratory ice data. It is publicly available on GitHub (see the data availability section at the end of the article). Given an IR absorption ice spectra in the range of 2.5--10.2 $\upmu$m (4000--980 cm$^{-1}$), it predicts the fractional composition of \ch{H2O}, \ch{CO}, \ch{CO2}, \ch{CH3OH}, \ch{NH3}, and \ch{CH4} (our target species) in the ice. Moreover, it also provides information about a representative temperature of the ice; although, as we discuss in Section 5, this can be an ill-defined parameter for astrophysical ices.

We note that for the analysis of the IR ice spectrum with AICE, the contribution from silicate absorption should first be removed. To do this, AICE is distributed along with a set of several other modules that allow us to pre-process the observed astronomical data: (i) merging and rebinning of spectra from different  telescopes and other instruments; (ii) continuum estimation and conversion to optical depth; and (iii) silicate removal (see Appendix \ref{appendix:pre-processing} for more details). In the following subsections, we describe how AICE works, providing details on its neural network architecture, its training, and the ice laboratory dataset used for the training.

\begin{figure*}
\centering
\includegraphics[trim={0 0 0 0}, width=0.8\hsize]{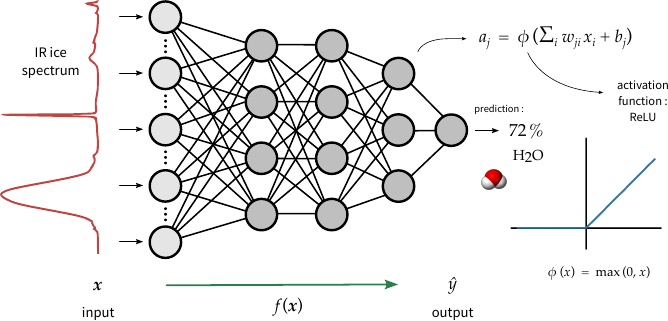}
\caption{Diagram of the model behind AICE: artificial neural networks. For each target molecule (\ch{H2O}, \ch{CO}, \ch{CO2}, \ch{CH3OH}, \ch{NH3}, and \ch{CH4}), a multilayer perceptron is used to predict the corresponding molecular fraction. The IR ice spectrum in absorbance is fed to the model, which transforms the  input  through a series of steps or layers. In each one, a linear combination of the previous values  ($x_i$) is  followed by the application of an activation function  ($\phi$),  obtaining $a_j = \phi \left (\sum_i w_{ji} \, x_i + b_j \right )$, with $w_{ji}$ and $b_j$ being the weights and bias of the layer $j$, respectively. We note that the actual neural networks used in AICE are larger than depicted in this scheme (see Section 2.2 and Fig. \ref{figure:aice-structure} for more details).}
\label{figure:aice-scheme}
\end{figure*}

\subsection{Artificial neural networks}

The mathematical model behind AICE is a set of artificial neural networks of the simplest kind: multilayer perceptrons. These models are composed of concatenations of linear combinations and non-linear activation functions that transform the input numeric vector in several steps (or layers). They yield a final output that can  either be a single value  or a vector \citep{Murtagh1991, Grosan2011}. Appendix \ref{appendix:neural-networks} contains a detailed description of multilayer perceptrons and how they work. 

In a nutshell, an artificial neural network is a mathematical function  with thousands of parameters, commonly known as weights \citep{LeCun2015}. Thanks to its mathematical structure and to a specific optimisation method, its parameters can be  successfully tuned to make the neural network mimic a particular  transformation, given enough pairs of inputs with their corresponding outputs \citep{Goodfellow2016, Cohen2020}. This process is called training, and is composed of several iterations (or epochs) where the difference between the predicted outputs and the actual outputs, is evaluated by using a certain error (or loss) function. During the training, thousands of pairs of input-and-output data are used to make the network learn the desired operation. In this procedure, a small fraction of the training dataset (e.g. $10\,\%$), is typically set aside and used in each epoch to evaluate the performance of the current model on data that has not been `seen' by the network. After  the training, we can apply the trained model to new input data to perform the referred operation or task and obtain a prediction of the output; this process is called inference.

\subsection{Architecture of AICE}

In the case of AICE, the input vector ($\boldsymbol{x}$) corresponds to the absorbance values of a spectrum given as a function of wavenumber from 4000 to 980 cm$^{-1}$, in steps of 1 cm$^{-1}$ (i.e.  3021 points). This input is normalised by its mean value of absorbance, which allows for a better generalisation; this normalisation is equivalent to a normalisation by area, but with the advantage of being dimensionless. 

The activation function used for the hidden layers is chosen to be the rectified linear unit (ReLU), $\phi(x) = \max(0,x)$. For the output, we want to obtain predictions for the fractional composition of our 6 target species in the ice, and of its temperature. Instead of using a single neural network to make all these predictions, we opted to use a different neural network for each parameter (i.e. one for each molecular compound, and an additional one for the ice temperature), since it yielded better results than just one neural network.  Therefore, we worked with a total of seven independent neural networks, one for the temperature and six for the molecular fraction of the six target species. Since the seven neural networks are independent, there is no direct dependence between the results of each neural network. If a dependence between temperature and molecular fractions were to be found, this would be an indirect effect inherent to the dataset (not an explicit dependence introduced by our neural networks). For each neural network, we inferred a single output value, $\hat y$. As the molecular fractional composition has to be a number between 0 and 1, we used a standard logistic sigmoid, $\phi(x) =  1 \, / \, (1 + {\rm e}^{-x})$, as the activation function in the output layer  for the six neural networks that predict the molecular fraction. In the case of the neural network that predicts the temperature, we just used a rectified linear unit (ReLU)  for the final output layer, $\phi(x) = \max(0,x)$, as the temperature has to be positive.

Figure \ref{figure:aice-scheme} shows a scheme of the general structure of the neural networks used in AICE, while Figure 2 shows the particular architecture employed. We  used three hidden layers, which in the case of a multilayer perceptron are called fully connected or dense layers. Their size (the number of neurons) starts with 120 for the first hidden layer and decreases progressively to 60 and 30,  ending with a final layer that returns a single value after applying the last activation function; in this way, we were able to decrease the number of activations and neurons in each layer from the 3021 points of the input to the single output. This is a common practice when building these kinds of models  (e.g. \citealt{Krizhevsky2012, Simonyan2014, He2016}). We  also used batch normalisation after each activation function, which yielded a total of 372,181 trainable parameters for each neural network (i.e. the weights and biases).  Dropout was applied only after the first hidden layer with a probability of 0.1.

With respect to the loss function, instead of using the typical mean squared error (MSE), we used the mean squared logarithmic error (MSLE). If $\hat y_k$ represents the predictions of the neural network for a certain parameter (fractional molecular composition or temperature) for the whole dataset (with size $m$) and $y_k$ represents the corresponding values registered from the experiments,  then the MSLE is defined as
\begin{equation}
    \indent \indent
    \mathcal{L} \; = \; \frac{1}{m} \, \sum_{k=1}^{m} \, ( \ln{(\hat y_k + 1)} - \ln{(y_k + 1)})^2  \;.
\end{equation}
The reason for using the MSLE instead of the MSE is to improve the accuracy of the predictions when the concentration of a molecule is very low or close to zero. Adding +1 in the logarithm avoids obtaining infinity when either the real or the predicted fraction of a molecule is exactly zero. In the case of temperature, we keep this loss function, as we also want to prioritise the accuracy of the model at low temperatures; for instance, an error of 1 K when the real temperature is 10 K is more important than when the real temperature is 70 K.

\begin{figure}
\centering
\includegraphics[trim={0 0 0 0}, width=0.7\hsize]{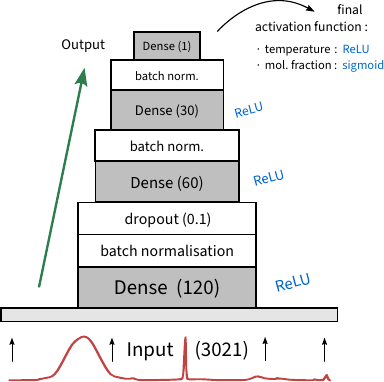}
\caption{Diagram of the architecture of the neural networks of AICE: Multi-layer perceptrons whose layer size (in parenthesis) decreases from the input spectrum to the output prediction. For the dropout step on the first dense layer, the probability of dropout is stated in parentheses.}
\label{figure:aice-structure}
\end{figure}

We  used the Adam method of gradient descent for training our model \citep{Kingma2014}, with an initial learning rate of $\alpha = 0.02$ and exponential decay rates of $\beta_1 = 0.9$ and $\beta_2 = 0.999$ (the two last values are typically used by default). The learning rate is reduced by multiplying it by 1/2 if there is no improvement in the validation loss after 15 epochs. The training is performed within a minimum of 80 epochs and a maximum of 160 epochs: within this range, it can be stopped whenever there is no improvement in the validation loss over the last 35 epochs, restoring the weights that yielded the lowest loss. To improve and fasten the training, we split our training subset in 12 batches (i.e. a batch size of 57 elements), and we apply batch normalisation after the activation of each hidden layer.  Appendix \ref{appendix:neural-networks} gives more details on the mathematics behind our networks. The aforementioned models were coded and trained in Python\footnote{\url{https://www.python.org/}} programming language and using the  Keras library, which is based on TensorFlow.\footnote{\url{https://www.tensorflow.org/}, \;\url{https://keras.io/}.}

\begin{figure*}
\centering
\includegraphics[trim={0 12 0 22}, clip, width=1.0\hsize]{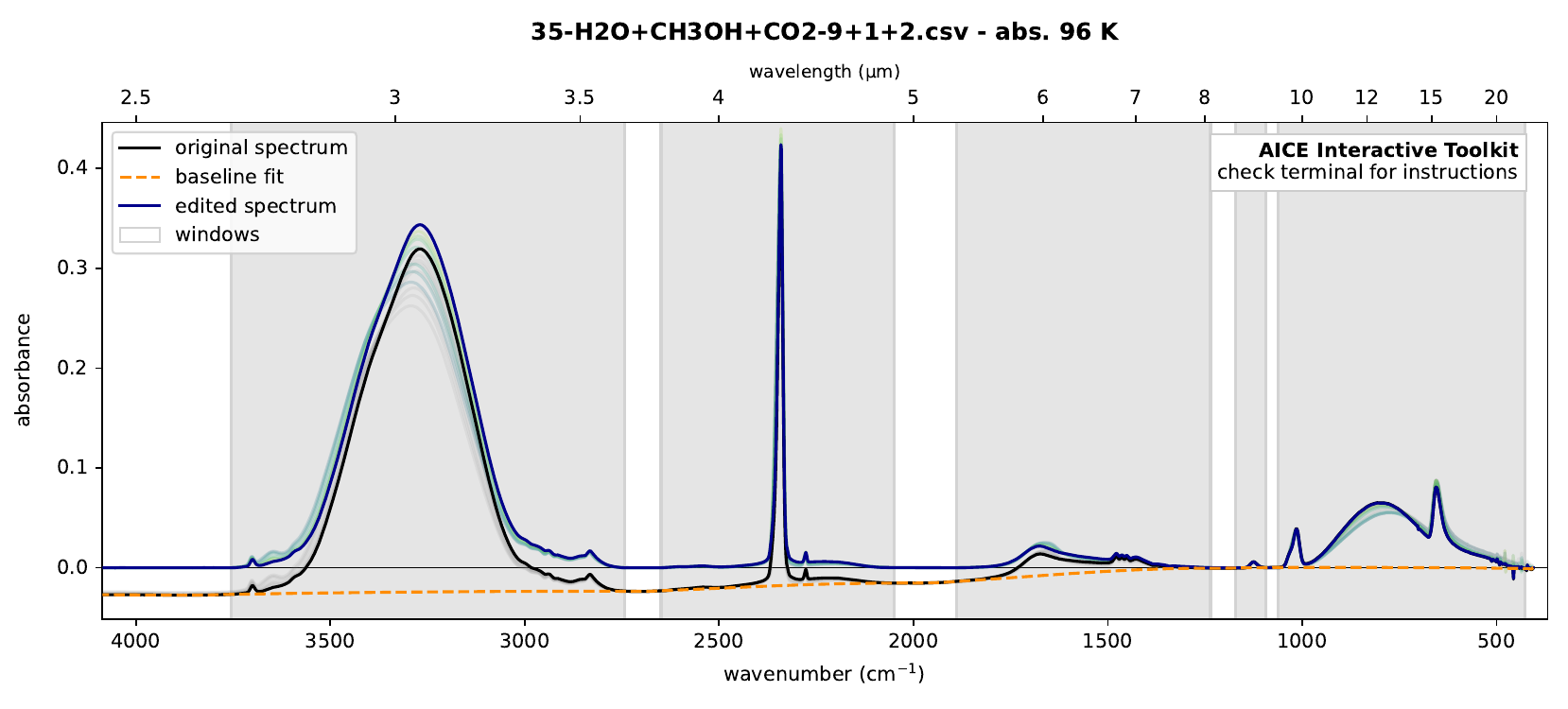}
\caption{Example of the baseline reduction for experiment \#35 from LIDA, performed with the interactive Python interface developed by us and included within AICE. The plot shows the spectra corresponding to this experiment (\ch{H2O}$\,:\,$\ch{CO2}$\,:\,$\ch{CH3OH}, 9:2:1), for several  temperatures. In black and blue, we display the original and reduced spectrum from the experiment at 10 K. In light grey and light green+yellow, we indicate the spectra for the rest of temperatures measured in the experiment up to 96 K. We  manually selected the windows that define the points of the continuum subtraction (white regions) and  the tool performed a baseline fit based on splines.}
\label{figure:aice-toolkit}
\end{figure*}

\subsection{Training dataset}

The predictions given by our tool AICE rely on the quality of the training dataset used to optimise the models. We  used a total of 571 laboratory IR spectra, with a varying chemical composition of the ice and temperature, both of which are known for each spectra. These data are the result of 212 different experiments with a certain ice composition, of which the majority correspond to different ice mixtures. The main ice components include \ch{H2O}, \ch{CO}, \ch{CO2}, \ch{CH3OH}, \ch{NH3}, and \ch{CH4} (our target species). However, note that we also included spectra from other mixtures with, for instance, formic acid (HCOOH) or acetaldehyde (\ch{CH3CHO}), as we also wanted our model to identify  cases where our target molecules are not present.

\subsubsection{Infrared molecular spectra}
\label{section:origin-data}

Most of the IR spectra used for the training (453 spectra, from 135 experiments) come from the Leiden Ice Database for Astrochemistry (LIDA) and are publicly available to visualise and download.\footnote{\url{https://icedb.strw.leidenuniv.nl/}} This database, presented in detail by \cite{Rocha2022}, now includes more than 1100 different spectra of more than 170 different ices, meaning that IR measurements for multiple temperatures are available. In most cases, the ice is formed at a low temperature (10--15 K) and the ice is then warmed up until it completely desorbs from the surface used. The majority of spectra are recorded in transmission mode; only a few are in reflection-absorption (RAIRS), with an angle of reflection of 45\degree.\footnote{This could lead to some inconsistencies for the same spectra due to differences in the light polarisation of the two methods. Despite this, given that we normalise the spectra,  the angle of reflection is 45\degree (larger differences would appear if light was almost parallel to the surface), and we have a rather small dataset, we still decided to  use those experiments.} These RAIRS spectra correspond to experiments \#\,55--69, mainly consisting of ices containing CO \citep{Fraser2004}. The IR range of the spectra in LIDA is approximately 4000--400 cm$^{-1}$ and the resolution varies from 0.25 to 2 cm$^{-1}$.
We did not use all the spectra available in LIDA, since we carried out a pre-selection and discarded some of them for different reasons, as explained below.
\vspace{-0.25cm}
\begin{enumerate}
\item We did not use the data from mixtures containing apolar molecules (such as \ch{O2} or \ch{N2}), since they do not present significant IR features, but only produce slight changes in the shape of the absorption features of other species contained in the mixture.
\item We did not use any processed ice (e.g. with UV radiation) or any ice where surface chemistry was involved, since in those cases we did not know the exact final chemical composition. 
\item Regarding the temperature, we also discarded any spectra in which any of the chemical compounds was clearly desorbed (i.e. their IR bands had disappeared with increasing temperature). This was the case for some ternary and quaternary mixtures, where it was evident that the most volatile species starts to desorb at relatively low temperatures and hence, the labelling of the composition is no longer correct. 
\item We discarded any spectra with a temperature greater than 100 K. One reason is to reduce the variability in the range of the parameters of the neural network to facilitate the training of the model, as the desorption temperature of the targeted species varies substantially; in any case, the only targeted molecules that desorb at higher temperatures than 100 K are water and methanol. Another reason is that for molecular clouds, starless or prestellar cores, and young protostellar systems, we expect to have temperatures quite lower than 100 K; although the temperature measured at the laboratory does not directly correlate with the temperature of the astrophysical environments (as we discuss in Section \ref{section:discussion}), we can expect that the differences will not be too large, so that we can model the ices contained in the mentioned types of sources with laboratory ices with temperatures lower than 100 K. 
\item We ruled out spectra from the same experiment with a difference in temperature of less than 8 K, as we wanted to ensure a significant difference in the shape of the measured spectra. However, we kept the spectrum with the highest recorded temperature for each experiment (not greater than 100 K) to ensure that all extreme cases were covered by the neural networks, within our desired temperature range.  By doing this, we tried to prevent the phenomenon of data leakage, where two very similar examples are present in both the training and validation subsets. As a result, the network can `memorise' the training example and achieve high performance in the validation example without actually learning to generalise, leading to an underestimation in the error. Therefore, data leakage can artificially inflate the predictions of the model and should be avoided \citep{Rosenblatt2024}.
\end{enumerate}
\vspace{-0.25cm}
The final number of spectra considered for our training dataset was thus 443 (i.e. making up almost half of the data stored in LIDA).

In addition to the LIDA spectra, we also used some spectra from other different databases and laboratories to complete our training dataset: 53 spectra from 19 different experiments from NASA's Optical Constants Database (OCdb)\footnote{\url{https://ocdb.smce.nasa.gov/}}; 30 spectra from 30 experiments from the database of the Universidade do Vale do Para\'iba (Univap)\footnote{\url{https://www1.univap.br/gaa/nkabs-database/data.htm}$\,$. The subsample of spectra we use here come from three different laboratories: 19 from GANIL (Grand Accelerateur National d’Ions Lourds, in France); 10 from LASA (Laborat\'orio de Astroqumica e Astrobiologia da Univap, in Brazil); and 1 from the Laboratory for Astrophysics of Leiden University (the Netherlands).}; 22 spectra from 20 experiments from NASA's Cosmic Ice Laboratory (CIL)\footnote{\tiny{\url{https://science.gsfc.nasa.gov/691/cosmicice/spectra.html}}}; 5 spectra from 5 different experiments from the Experimental Astrophysics Laboratory (Laboratorio di Astrofisica Sperimentale, LASp) in the Astrophysical Observatory of Catania (INAF)\footnote{\url{https://oldwww.oact.inaf.it/weboac/labsp/}}; 3  spectra from the same single experiment from the National Synchrotron Radiation Research Center (NSRRC)\footnote{\url{https://www.nsrrc.org.tw/english/index.aspx}} in Taiwan; 2 spectra from the same experiment from the Institute of the Structure of Matter (Instituto de Estructura de la Materia, IEM; CSIC)\footnote{\url{https://www.iem.csic.es/fismol/ices/index_en.shtml}} in Spain; and, finally, 2 spectra from 2 different experiments presented in the work by \cite{McClure2023}. All of them were recorded in transmission and have similar experimental properties to the ones from LIDA. Table 1 on the \href{https://zenodo.org/records/16902313}{supplementary data} provides a complete list of all the spectra collected for this work.

\subsubsection{Spectra processing and baseline removal}

All of these 571 IR spectra had to be processed and reduced before using them to train our neural networks.
First, we checked whether the spectra had been reduced or they had
baseline issues that could distort the spectrum even in regions with only
continuum.
 In cases where the baseline issues were evident, we subtracted these baselines from our data by using an interactive Python tool we developed and included within AICE. This tool allows the user to manually select the continuum regions of the spectra that will be used to fit the baseline, and to subtract it. The baseline fit is done with splines, namely, a concatenation of polynomials of third order, where the smoothness of the resulting fit can be adjusted. In Fig. \ref{figure:aice-toolkit}, we give  an example of this procedure. We refer to Appendix \ref{appendix:aice-toolkit} for more details on the method.

Another point that had to be taken into account was the presence of absorption bands from residual compounds that were not listed in the composition of the experiment, for instance, solid water present on the IR detector, or gaseous \ch{CO2}  or even \ch{H2O} present in the chamber of the sample. In addition, curve deformations that are not linked to any absorption signatures can come from experimental artifacts and induce a bias in the recognition of features by AICE. We removed these features by interpolating the spectra with polynomials (splines). Similarly, there can be signals of molecular isotopologes due to the impurity of the experimental samples, like the \ch{^{13}C} variants of \ch{CO} and \ch{CO2}; this can indeed be seen in Fig. 3 for \ch{^{13}CO2}, with the small peak next to the \ch{CO2} absorption line at $\sim$2340 cm$^{-1}$. We removed  most of these signals of \ch{^{13}C} isotopologues but we left  them for some of the spectra, so that our models are trained with and without the signal and learn to ignore it when determining the presence of the \ch{^{12}C} isotopologues. Lastly, some spectra presented regions that were too noisy. Hence, we  smoothed those parts to obtain spectra with uniform and lower levels of noise. We did not smooth any thin feature or line, as that would deform the shape of the spectrum. An example of the baseline subtraction, artifacts removal and spectra smoothing is shown in Appendix \ref{appendix:aice-toolkit}, along with more detailed explanations. Finally, we  stress that the results predicted by AICE are sensitive to the shape of the spectrum (especially for temperature); hence, it is essential to have well-reduced laboratory spectra to achieve the best performance of AICE.

\begin{figure*}
\centering
\includegraphics[trim={0 10 0 30}, width=0.78\hsize, clip]{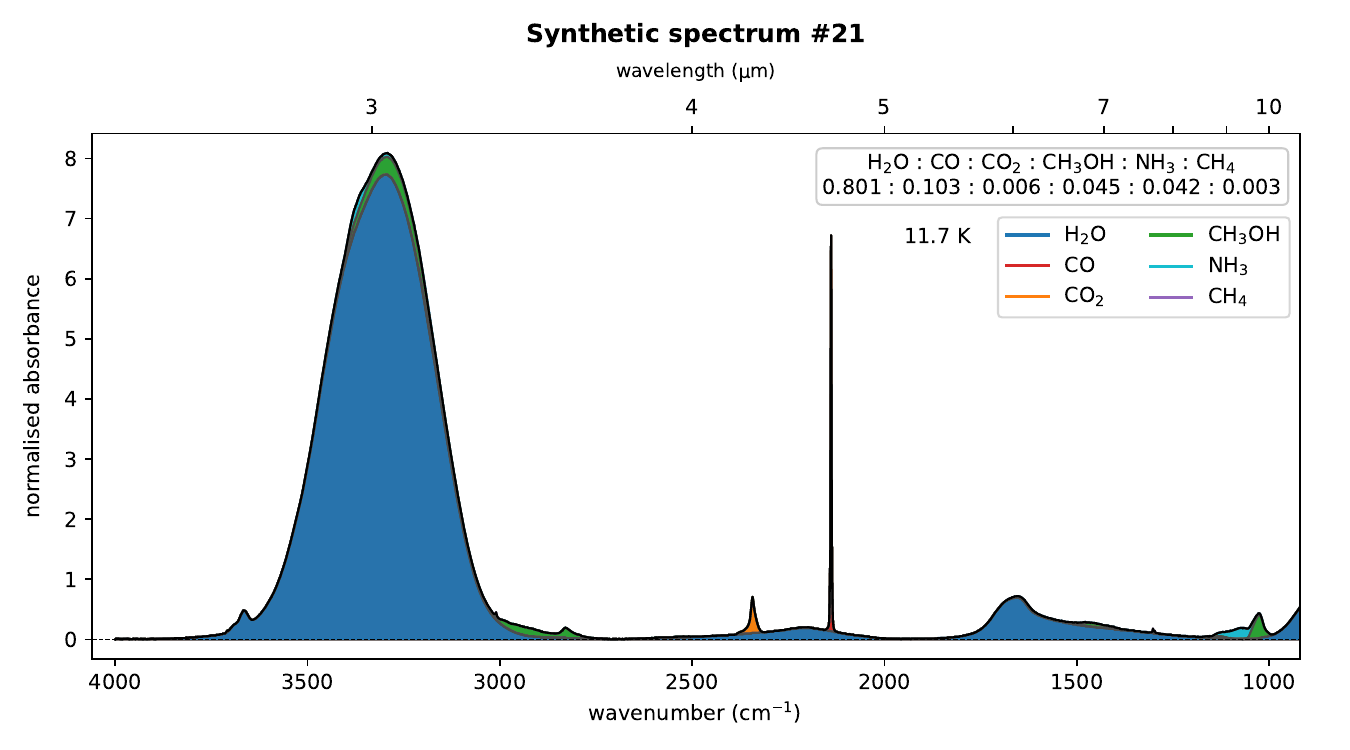}
\caption{Example of synthetic generated spectrum (\#21), normalised by its mean absorbance value. The colours represent the different molecular contributions. This spectrum represents a layered mixture of \ch{H2O}, CO, \ch{CH3OH}, \ch{NH3}, \ch{CO2}, and \ch{CH4}, in decreasing order molecular fraction, all of them at a temperature of 11.7 K. The pure ice spectra used to construct this curve come from different experiments from LIDA, OCdb and Univap (see Table \ref{table:pure-ices} in Appendix \ref{appendix:coefficients}), at different temperatures close to 11.7 K (see Section \ref{section:data-agumentation} for more details).}
\label{figure:synthetic-spectrum}
\end{figure*}

\subsubsection{Resampling and normalisation}

As mentioned before, AICE works with a range of wavenumber values from 4000 to 980 cm$^{-1}$ with a separation of 1 cm$^{-1}$ (3021 points). As not all the spectra have this same spacing, we resampled them to a resolution of 1 cm$^{-1}$. In the cases where the resolution was worse than 1 cm$^{-1}$, we supersampled the spectra; for instance, for a resolution of 2 cm$^{-1}$, we just doubled the number of points by linearly interpolating between each pair of values. Although supersampling is not a univocal transformation, in most cases there were no thin absorption bands; therefore we did not introduce any artificial deformation in the spectra. We note that, from our selected spectra, most of the ones corresponding to ices containing CO (the species with the thinnest and sharper absorption band) have a resolution of at least 1 cm$^{-1}$; in particular, only nine spectra from three experiments have a poorer resolution (2 cm$^{-1}$ in all cases): LIDA \#124, OCdb \#\,23--26, and OCdb \#\,29--32. Besides, these spectra correspond to ice mixtures, and normally the CO band gets naturally broadened when mixed with other molecules. Therefore, our supersampling to 1 cm$^{-1}$ should not entail significant deformations of the real shape of the CO band.

After resampling, we normalised the spectra simply dividing by the mean of the absorbance values within the AICE range. This is equivalent to normalising by the area, as the integrated area of a set of points is proportional to its mean, given a fixed range in wavenumber. In this way, we generalise our data to any combination proportional to the original ice mixture.

\subsubsection{Data augmentation using pure ice spectra}
\label{section:data-agumentation}

Out of our six targeted species, ammonia and methane present very few experiments in our dataset, which makes the training of our networks a challenging task. In order to tackle this problem, we opted to enlarge our dataset by generating synthetic mixtures using linear combinations of experimental pure ice spectra. To do this, we employed pure ice spectra from LIDA, OCdb and Univap for our six target species.

For each pure spectrum, we integrated one of the strongest bands, and used the experimental band strengths to derive the column density of the ice. Dividing the absorbance by the column density we obtain the absorption cross-section of each pure spectrum. The cross-sections of different species can then be summed up after weighing each contribution by the column density of each molecule. However, since we used the mean as normalisation, we did not need to choose any column density for each species; we simply used the relative proportions of the desired ice mixture. Afterwards, we divided the final spectrum by its mean (see Fig. \ref{figure:synthetic-spectrum} for an example of synthetic spectrum). This process can be repeated for the different temperatures measured for each pure species, obtaining a set of normalised cross-sections for each temperature.

In reality, the spectra of ice mixtures are not just a linear combination of the pure ice spectra, as there are interactions between molecules that deform the shape, the position  and even the intensity of the bands. However, the absorption spectra of a combination of separated layers of different species is actually the linear combination of the pure ice spectra (apart from a tiny contribution of the interfaces, which is insignificant in our dataset, with hundreds of monolayers per ice). Therefore, these synthetic combinations from pure ices can be seen both as a first approximation to mixed ice spectra and as the spectra corresponding to combinations of ice thick layers with different molecules. Actually, there are some experiments in LIDA that correspond to two superimposed layers of pure ices at the same temperature.

By using the method described below, we generated a total of 282 synthetic spectra containing different ice mixtures of our 6 targeted species. We chose the values of each molecular proportion within the mixture using a random algorithm that ensures that all of the species are represented in a balanced way, as explained in Appendix \ref{appendix:coefficients}. Regarding the temperatures, we note that species such as CO and CH$_4$ desorb at temperatures $\gtrsim$25 K,  and this to be taken into account for creating the augmented dataset, as we explain in the following paragraph. In addition, our pure NH$_3$ data only has measured laboratory spectra up to 70 K. Therefore, the temperature range for data augmentation
needs to reflect these limitations. 

Thus, we employed the following procedure. For each synthetic mixture, when all species present a concentration higher than 2$\,\%$, we restricted the allowed temperature range to the common range of possible temperatures  according to our pure ice spectra (i.e. between 10 K and 25 K). However, if in the synthetic mixture one of the compounds presented a concentration lower than 2$\,\%$, we ignored the maximum temperature limitation for that species and used the pure ice spectrum at the maximum temperature available if the temperature of the synthetic mixture was greater than that. This helps the neural networks to learn the behavior of high-temperature ices for major molecular compounds. In any case, we note that species such as CO and CH$_4$, which desorb in large amounts at temperatures $\gtrsim$30 K, could remain in the ice in small quantities at higher temperatures when mixed  \citep{Simon2019, Kruczkiewicz2024}.  Once the temperature range was defined, we randomly selected one temperature within that range, with a uniform probability. We then performed a linear interpolation of the pure ice spectra of each molecule in the dimension of temperature, to obtain the spectrum of each pure ice at the selected temperature. The final synthetic spectrum was built through the linear combination of the interpolated pure ice spectra using the proportions previously determined for this particular mixture.

Therefore, we obtained a complete dataset of 853 spectra (571+282), which was used to train our neural networks. We stress that a larger number of synthetic spectra in our dataset would spoil the performance of the networks since the synthetic spectra do not account for the molecular interactions in the ice mixture that get reflected in the real molecular band profiles. With our procedure, the  synthetic spectra represent two-thirds of the total dataset, providing an optimum performance for the networks. 

\subsubsection{Training and validation subsets}

To train our neural networks, we had to split our final dataset of 571+282 spectra into a training and a validation subset. 
To do this, we used the following procedure:
\vspace{-0.25cm}
\begin{itemize}[leftmargin=12pt]
    \item First, we used all of the pure ice spectra for training, as we want to ensure that our models can recognise the simplest case of a pure mixture. This applies to all pure ices, including the species not targeted by our model. By doing so, we included 141 spectra in our training subset.
    \item Then, we used for training any spectrum containing mo-lecules different from the target species, even if the concentration is small. This was done so that the networks could learn when the target species were not present in the ice. This added another 198 spectra to our training subset.
    \item For the rest of the laboratory spectra (233) and the 282 synthetic spectra, we randomly shuffled and split them so that the validation subset constituted the 20\,\% of the whole dataset.
\end{itemize}
\vspace{-0.25cm}
Following this procedure, we ended up with a total of 681 spectra for training and 171 for validation. However, this number is rather low for an optimal trainingof  artificial neural networks with hundreds of thousands of parameters. Furthermore, the data that have been selected to be in the validation subset would not be used to fit the parameters of our models, so we  lose their information for the training of the networks.

To address this, we used a method called bagging  \citep{Breiman1996}, with a ten-fold cross-validation sampling scheme \citep{Petersen2007}. We made ten different shuffles of our data with fixed random seeds,\footnote{The random seed is an integer number that defines the sequence of random numbers generated by the computer.} so that we ended up with ten different pairs of training and validation subsets. Then, we trained our neural networks for each of them, obtaining different values for their parameters at the end of the training. Finally, we used all of the ten variants of our models to obtain a prediction for the input spectra, being the result the mean of the output results. In this way, for each of our target variables we ended up with an ensemble of neural networks that not only accounted for the different shuffles (or splits) of our training dataset, but also allowed us to explore the different initialisations of the parameters of the neural networks, since they are randomly set at the beginning of the training \citep{Sagi2018}.  The main disadvantage of this method is that it increases the computation time of inference (by a factor of 10 in our case). In any case, our models are so fast in terms of inference (as noted in Section \ref{section:aice-astro-test}) that this is not an issue.

To summarise, we trained seven sets of ten neural networks: one set for each targeted variable (temperature and molecular fraction for our six target species). Each set of models contains ten variants that are trained with the different shuffles of our whole dataset of  571+282 spectra. The final AICE result is the mean of the output results of the ten neural networks, for each of the seven target variables. Additionally, for each prediction, we computed the standard deviation between the ten model variants, which were considered as the uncertainty in the AICE prediction.

\subsection{Choosing the model hyperparameters}

The parameters that define the architecture of the neural network, such as the number of layers and number of neurons per layer, as well as the choice of the loss function (MSE, MSLE...) and training method  constitute the hyperparameters of the neural network. More examples of them are the batch size and the learning rate.

The choice of the exact values of these parameters starts with an educated guess. In our case, we performed several tests varying some of the values and checking the performance of the models after the training procedure, using the validation dataset for it. This has to be done with several random seeds (thus, different splits of training and validation) to reduce the possibility of biasing our hyperparameters to a specific seed. For example, we first tried with an additional hidden layer with 480 neurons, although after some tests, we realised that we could work without this layer, obtaining the same performance with the benefit of reducing the model size by a factor of 4. We also tried using 1D convolutional neural networks, a type of models which is commonly used for pattern recognition \citep{LeCun2015}. However, the final performance was not as good as the one obtained with our multi-layer perceptron and, thus, we opted for this simple yet powerful type of model.

In a similar way, we also tried using a unique neural network with several final outputs, one for each of the predicted variables (six molecular fractions + ice temperature). However, we could not reach the same performance as the one achieved using individual neural networks, probably due to the inhomogeneity of the training dataset with respect to our parameter space (temperature and molecular fractions). Focussing only on predicting one variable in each neural network reduces the difficulty of the optimisation task, yielding a better individual accuracy for each model. Besides, this modular approach should facilitate the extension of our model with additional targeted species in the future, since we would only need to train those additional networks, without retraining the models presented in this work. Lastly, similar tests were performed when choosing the number of synthetic spectra to be included in the augmented dataset and the minimum separation in temperature in the original laboratory dataset. 

\section{Results}

In this section, we present the results of the model training and report the validation of the networks. After these steps, we  tested AICE on two ice absorption spectra recently obtained with the JWST towards the background stars NIR38 and J110621 in the Chamaeleon I molecular cloud \citep[reported by][]{McClure2023}. 

\subsection{Training and validation}
\label{section:training}

Training each neural networks takes approximately 21\,s, so training each set of seven neural networks takes about 150\,s.\footnote{Processor: Intel i5, 4 cores, 2 GHz; RAM: 16 GB, DDR4, 3.7 GHz.} Repeating this process ten times, for ten different random seeds (from 1 to 10), we completed the training of our model AICE in around 25 min.

\begin{figure*}
\centering
\includegraphics[trim={0 15 0 40}, clip, width=\hsize]{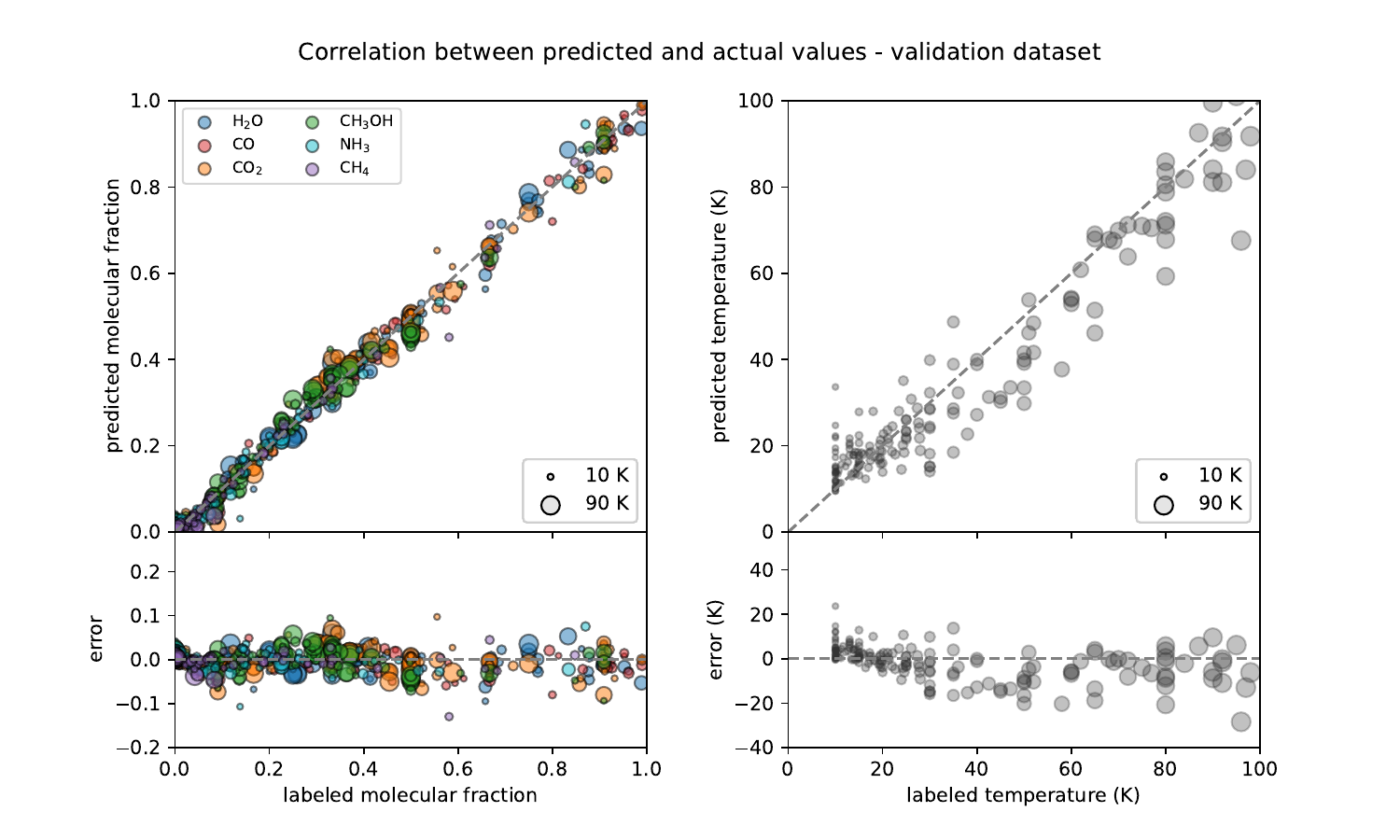}
\caption{Performance of our neural networks after the training with the first training+validation split, with respect to the spectra included in the validation subset. The plots show the predicted versus true labeled values for the molecular fractions and temperature. {\it Left panels}: Molecular fractions of ice spectra. Each ice spectrum in the validation subset (with a certain composition and temperature) corresponds to six circles in the plot: one for each targeted molecule (denoted by colours). The sizes of the circles indicate the temperature of the ice. The bottom panel reports the difference between the predicted molecular fraction and the labeled value from the experiment. {\it Right panels}: Temperature of ice spectra, represented by each circle. Again, the bottom panel reports the difference between the predicted and the labeled values.}
\label{figure:aice-validation}
\end{figure*}

Figure \ref{figure:aice-training} in Appendix \ref{appendix:aice-training} shows the evolution of training and validation losses during the training for a certain seed, while Fig. \ref{figure:aice-validation} shows the results of the validation for the same seed. These validation plots constitute a powerful way of determining how well the networks have learnt to generalise, since their performance is assessed for data that were not used in the optimisation process. In these plots, we compare the predicted molecular fractional composition and temperature with the real ones from the labeled spectra of the validation subset. As it can be seen in the upper panels, all points follow the 1:1 line (with a certain dispersion), which indicates that the networks have properly learnt to generalise, at least in the parameter space present in our training dataset. The observed dispersion can be due to two reasons: possible experimental systematic effects in the laboratory spectra used for the training, and the ability of the models to learn from the training dataset. In any case, the total dispersion is a constraint on these two possible effects and, as observed in Figure 5, the dispersion is relatively low. In addition, we note that Fig. \ref{figure:aice-validation} shows the validation results for both real laboratory spectra and the synthetic mixtures from pure ices, and both populations of points follow the 1:1 line. As it can be seen in the lower panels, the resulting error for the molecular fraction for the different target species typically lies below 0.1 ($10\,\%$), while it is usually below 20 K for temperatures. We note that the temperature tends to be underestimated for ices with real (labeled) temperatures $\gtrsim$30 K, which is something that, in a lower or greater extent, we observed for all the 10 dataset splits. This could be due to the fact that most of the spectra in our dataset correspond to ices with temperatures $\lesssim$30 K.

\begin{table}
\caption{RMSE obtained in the validation of the neural networks after training, for each targeted variable and each random split of the dataset.}
\label{table:aice-errors}
\centering
\begin{tabular}{cccccccc}
\hline
 & \multicolumn{7}{c}{\bf root mean squared error (RMSE) in validation} \tabularnewline
 \cline{2-8}
{\bf split} & \multicolumn{6}{c}{{\bf fractional composition} (\%)} & {\bf temp.} \tabularnewline
\# & \ch{H2O} & \ch{CO} & \ch{CO2} & \ch{CH3OH} & \ch{NH3} & \ch{CH4} & (K) \tabularnewline
\hline 
1 & 1.9 & 1.6 & 2.5 & 2.4 & 1.6 & 1.4 & 7.7 \tabularnewline 
2 & 2.5 & 2.2 & 3.3 & 2.1 & 1.1 & 1.6 & 12.6 \tabularnewline 
3 & 2.2 & 2.6 & 2.1 & 2.3 & 0.9 & 1.1 & 9.6 \tabularnewline 
4 & 2.8 & 2.5 & 2.3 & 2.0 & 1.4 & 1.2 & 9.9 \tabularnewline 
5 & 2.0 & 2.7 & 2.3 & 2.3 & 1.0 & 1.4 & 10.5 \tabularnewline 
6 & 2.4 & 2.3 & 2.3 & 2.6 & 1.1 & 1.7 & 9.1 \tabularnewline 
7 & 2.6 & 1.9 & 2.4 & 2.4 & 1.2 & 1.5 & 11.7 \tabularnewline 
8 & 2.6 & 1.7 & 2.7 & 1.9 & 1.6 & 1.4 & 10.2 \tabularnewline 
9 & 2.9 & 3.3 & 2.3 & 2.1 & 2.1 & 1.5 & 9.2 \tabularnewline 
10 & 2.4 & 2.3 & 2.8 & 2.2 & 1.4 & 1.3 & 11.0 \tabularnewline
\hline
\tiny{mean} & 2.4 & 2.3 & 2.5 & 2.2 & 1.4 & 1.4 & 10.2 \tabularnewline 
\hline 
\end{tabular}
\end{table}

To be more precise with respect to the errors, we used the root mean squared error (RMSE) to estimate the accuracy of the neural networks in their predictions. The list of root mean squared errors in the validation subset for the different individual neural networks and random seeds (and therefore splits of the dataset in training+validation) is shown in Table \ref{table:aice-errors}.  The RMSE in the fractional composition, after averaging for the six species and for the ten splits, is 0.021 (2.1\,\%); whereas for the temperature, we have an average RMSE of 10.8 K. This validation error can be a slight underestimation of the actual model error, since the validation subset indirectly affects the evolution of the training (for example, reducing the learning rate). Hence, to have a more unbiased estimation of the model error, we used the nested cross-validation method \citep{Cawley2010}. In this algorithm, we perform several iterations (also called trials; here, we chose five) in which we divide our original training dataset (571+282 spectra) in two parts: a training+validation subset (90$\,$\%) and a test subset (10$\,$\%). Then, for each of the training+validation subsets, we perform the full training of our 7 neural networks by using ten random seeds, which define new splits of the training+validation subset in training subset and validation subset, using proportions of 80/20. For each of these 10 last splits, we evaluate the error of the model on the corresponding test subset, obtaining in total 10 different values of this error. For each of the five initial iterations (or trials, which define a different test subset), we generated ten new values for the model error.  After all these calculations, averaging all the errors (50 in total) we obtain a value of 2.7 \% for the molecular fraction and  12.8 K for the temperature, which are a bit higher than the validation errors mentioned in the previous paragraph; for each molecule, the errors are: 2.5 \% for \ch{H2O}; 3.1 \% for CO; 3.0 \% for \ch{CO2}; 2.6 \% for \ch{CH3OH}, 1.7 \% for \ch{NH3}; and  2.4 \% for \ch{CH4}. Those values can be considered as upper limits on the actual model errors, since in these trials the models are trained with a fraction of the whole training dataset and hence their performance is expected to be worse. The actual average error of the AICE models would be in between the average validation error and the average test error obtained with nested cross-validation.

We also need to bear in mind that one potential source of error comes from uncertainties in the labelling of the laboratory spectra.  While the temperature is a very well determined parameter, the determination of the exact composition is not trivial. Sometimes, premade gaseous mixtures can be used in the experimental chamber, although the different molecular velocities can lead to a slightly different ice composition. Other times, the IR molecular bands can be integrated and scaled by band strengths from pure ices, but this can lead to errors up to $\sim$20\,\%. A more precise method involves knowing the input pressures of each gas in the chamber and carefully calibrating the fluxes and pressures of each species. In addition, we have to bear in mind that different experimental set-ups can have different systematic effects that change the shape of the recorded spectrum.  Therefore, these sources of uncertainty can be transferred to the training of our models, which translates into a higher dispersion of the predictions.  These effects can be constrained in the validation step, since the observed error can be due either to those systematic errors, to errors in the labelling of the composition, or to the performance of the models itself. As a clear example, we highlight the cluster of circles at a molecular fraction of 0.5 (50$\,$\%) in the left panel of Figure \ref{figure:aice-validation}, since many spectra are binary mixtures with ratios 1:1. To explain the  larger dispersion for some of the  molecular fractions, we suspect that their labeled composition is somewhat uncertain given the aforementioned possible experimental systematic effects, although it could simply happen that the neural network is yielding a larger error than usual.\footnote{During some of the initial trainings of our models, we identified prominent outliers, which turned out to be mislabeled spectra; after correcting the labelling, those points were no longer outliers, showing typical errors of less than 0.03 (3$\,$\%). This demonstrates the utility of these models to properly identify the ice composition.}  Regarding the temperature, it is indeed a delicate variable, since in the experiments used to train AICE, the lower temperature corresponds to the temperature at which the ice was initially deposited, while the following temperatures in increasing order correspond to the annealing of the cold deposited ice. However, the effects of this annealing in the shape, position and intensity of the molecular bands, can be also obtained experimentally by other means, such as by the type of deposition, as discussed in Section \ref{section:discussion}.

\begin{figure*}
\centering
\includegraphics[trim={0 18 0 10}, clip, width=\hsize]{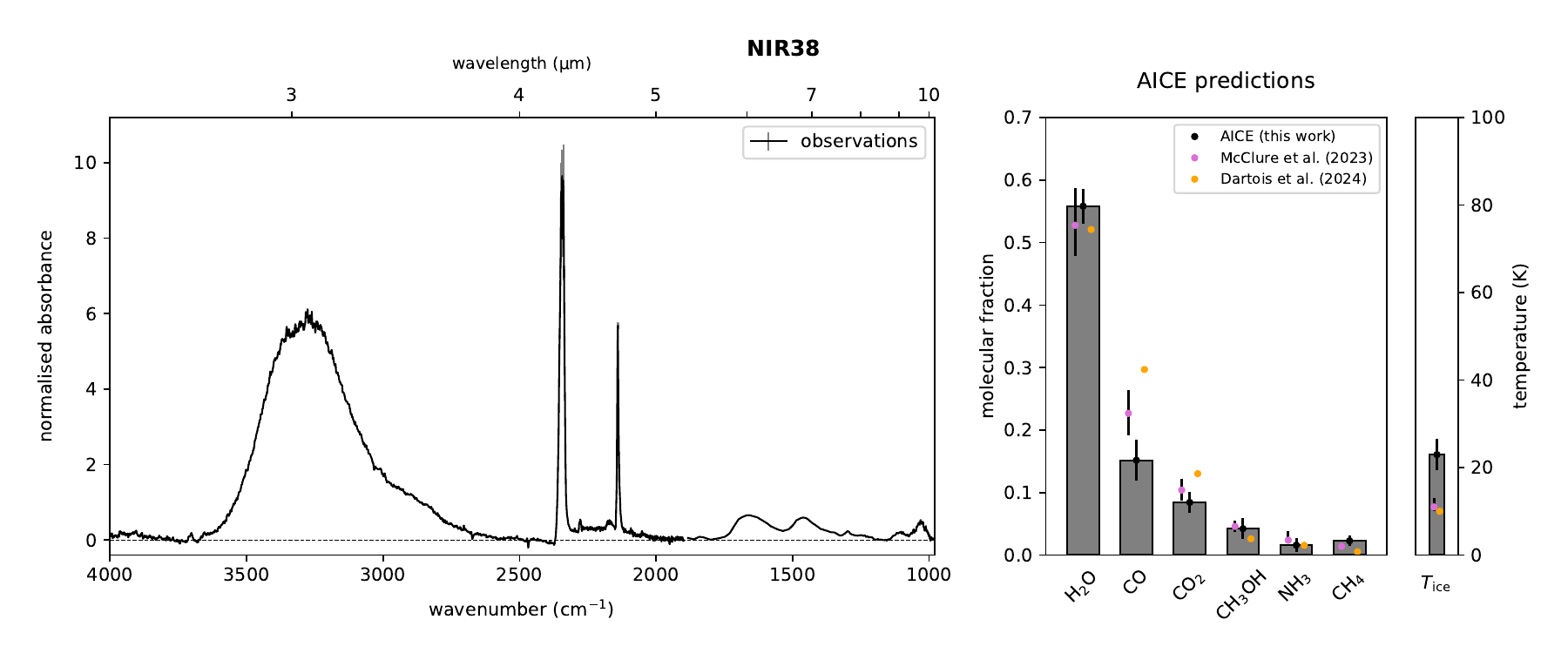}
\includegraphics[trim={0 10 0 10}, clip, width=\hsize]{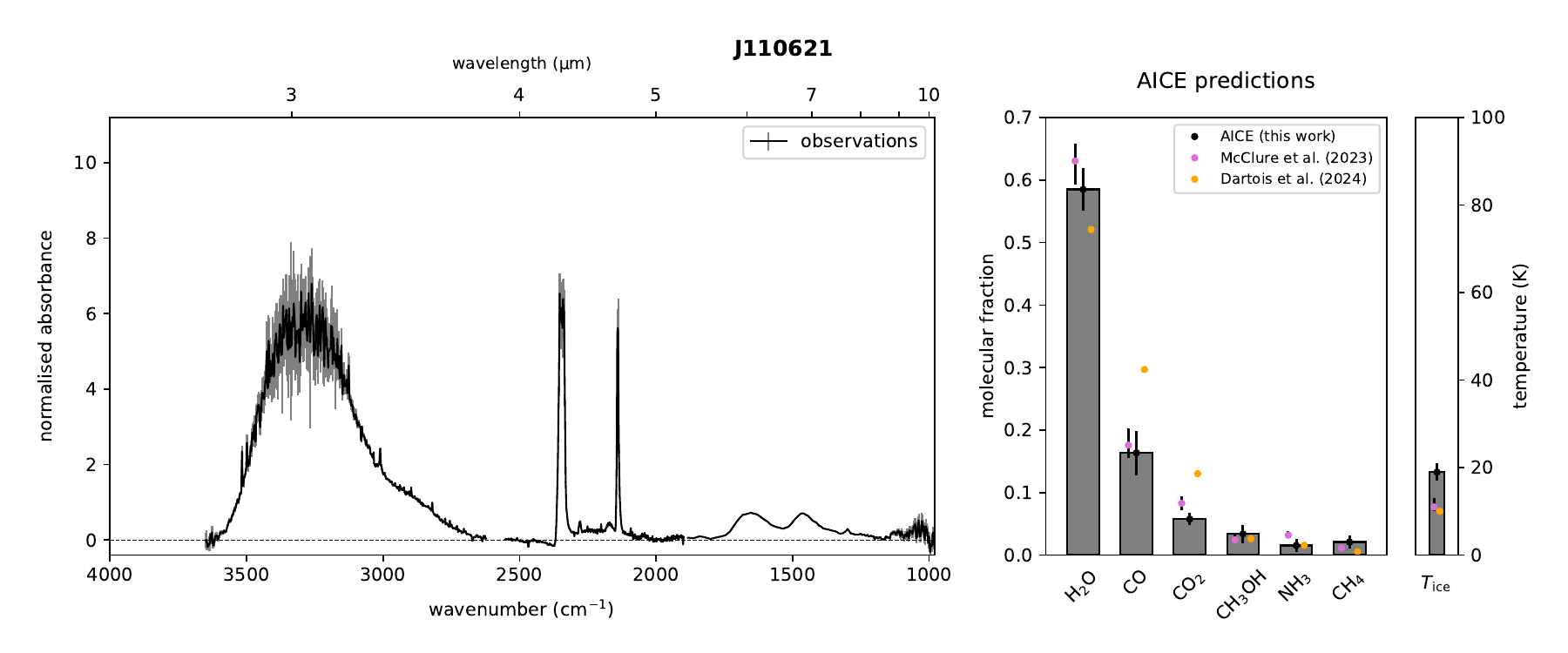}
\caption{Absorbance spectra obtained by JWST towards the background stars NIR38 (top) and J110621 (bottom) in the Chamaeleon I molecular cloud, and predictions obtained by AICE for the corresponding ice composition and temperature. {\it Left panels:} Absorbance spectra (in black) obtained after the reduction of the original JWST spectra in flux \citep{McClure2023}. {\it Right panels:} Direct numerical predictions given by AICE of the fractional molecular composition and temperature of the corresponding ices towards both positions (grey bars and black points), compared to estimates obtained by other authors (in pink, \citealp{McClure2023}; in orange, \citealp{Dartois2024}).}
\label{figure:aice-test}
\end{figure*}

As mentioned in the previous section, the final result of AICE will be, for each targeted variable (molecular fractions and temperature), the mean value of the predictions of the 10 variants corresponding to each random seed. Similarly, the uncertainty in the prediction will be the standard deviation of the predictions of those 10 variants, for each targeted variable. In principle, the variance between the different versions of the models will be related to the observed dispersion in the validation step (Fig. \ref{figure:aice-validation} and Table \ref{table:aice-errors}), which are related to the possible systematic effects from laboratory experiments (see above). Therefore, the effect of those possible systematic errors would get transferred to the variance between the models predictions, and hence to the uncertainty in the AICE output.

\subsection{Testing AICE with JWST data}
\label{section:aice-astro-test}

\renewcommand*{\arraystretch}{1.4}
\begin{table*}
\caption{Comparison between the AICE predictions with the results of \citet{McClure2023} and \citet{Dartois2024} for NIR38 and J110621.}
\label{table:aice-test}
\centering
\begin{tabular}{lccccc}
\hline 
 & \multicolumn{2}{c}{\textbf{NIR38}} &  & \multicolumn{2}{c}{\textbf{J110621}}\tabularnewline
\cline{2-6}
\multirow{2}{*}{\textbf{prediction}} & \textbf{AICE} & \textbf{McClure} & \textbf{Dartois} & \textbf{McClure} & \textbf{AICE}\tabularnewline
 & \textbf{(this work)} & \textbf{et al. (2023)} & \textbf{et al. (2024)} & \textbf{et al. (2023)} & \textbf{(this work)}\tabularnewline
\hline 
{\bf $\mathbf{H_{2}O}$} (\%) & $56 \pm 3$ & $53_{-5}^{+6}$ & 52 & $63_{-4}^{+3}$ & $59 \pm 3$ \tabularnewline 
{\bf CO} (\%) & $15 \pm 3$ & $23_{-4}^{+3}$ & 30 & $18_{-2}^{+3}$ & $16 \pm 4$ \tabularnewline 
{\bf $\mathbf{CO_2}$} (\%) & $8.4 \pm 1.7$ & $10.4_{-1.7}^{+1.8}$ & 13 & $8.3_{-1.1}^{+1.2}$ & $5.8 \pm 1.0$ \tabularnewline 
{\bf $\mathbf{CH_{3}OH}$} (\%) & $4.2 \pm 1.7$ & $4.6_{-0.9}^{+1.0}$ & 2.6 & $2.4_{-0.5}^{+0.9}$ & $3.4 \pm 1.5$ \tabularnewline 
{\bf $\mathbf{NH_{3}}$} (\%) & $1.6 \pm 1.1$ & $2.4_{-0.4}^{+1.5}$ & 1.6 & $3.2_{-0.4}^{+0.7}$ & $1.5 \pm 1.1$ \tabularnewline 
{\bf $\mathbf{CH_{4}}$} (\%) & $2.3 \pm 0.8$ & $1.38_{-0.18}^{+0.20}$ & 0.52 & $1.16_{-0.13}^{+0.13}$ & $2.1 \pm 1.0$ \tabularnewline 
{\bf temperature} (K) & $23 \pm 4$ & $11_{-1}^{+2}$ & 10 & $11_{-1}^{+2}$ & $19.0 \pm 2.0$ \tabularnewline 
{\bf CO\,$/$\,$\mathbf{H_{2}O}$} (\%) & $27 \pm 6$ & $44_{-7}^{+9}$ & 57 & $29_{-4}^{+5}$ & $27 \pm 7$ \tabularnewline 
{\bf $\mathbf{CO_2}$\,$/$\,$\mathbf{H_{2}O}$} (\%) & $15 \pm 3$ & $19 \pm 4$ & 25 & $13.3_{-1.9}^{+2.0}$ & $9.8 \pm 1.8$ \tabularnewline 
{\bf $\mathbf{CH_{3}OH}$\,$/$\,$\mathbf{H_{2}O}$} (\%) & $8 \pm 3$ & $8.7_{-1.9}^{+2.2}$ & 5 & $3.8_{-0.8}^{+1.4}$ & $5.8 \pm 2.5$ \tabularnewline 
{\bf $\mathbf{NH_{3}}$\,$/$\,$\mathbf{H_{2}O}$} (\%) & $2.9 \pm 2.0$ & $5_{-1}^{+3}$ & 3 & $5.1_{-0.7}^{+1.1}$ & $2.5 \pm 1.9$ \tabularnewline 
{\bf $\mathbf{CH_{4}}$\,$/$\,$\mathbf{H_{2}O}$} (\%) & $4.1 \pm 1.5$ & $2.6_{-0.4}^{+0.5}$ & 1 & $1.85_{-0.22}^{+0.24}$ & $3.6 \pm 1.7$ \tabularnewline 
\hline 
\end{tabular}
\begin{justify}
\footnotesize{{\bf Note.} Individual molecular fractions are given with respect to the total quantity of all the predicted species; in the case of the two other works, we calculated these fractions with the data available in the corresponding papers. For the temperature, \cite{McClure2023} do not present a unique temperature, but several possible ones ranging from 10 to 16 K; consequently, we adopted the value of $11_{-1}^{+2}$ K as an estimate of a representative temperature. The estimates by \cite{Dartois2024} are the same for both background stars (i.e. the one of the laboratory spectrum used in their radiative transfer calculations), and thus these values are placed at the centre of the table. The fractions with respect to water are derived from the direct results of AICE and those reported by the cited authors, using the Python library RichValues (see Appendix \ref{appendix:richvalues} for more details).}
\end{justify}
\end{table*}
\renewcommand*{\arraystretch}{1.0}

Once we have trained and validated our model, we can test how it performs with astronomical data. To do this, we use the JWST spectra reported by \cite{McClure2023} obtained within the IceAge ERS project. It consists of IR ice absorption spectra measured towards two background stars in the Chamaeleon I molecular cloud: NIR38 and J110621.\footnote{The original spectra are available in Zenodo: \\ \url{https://zenodo.org/records/7501239}} The spectra were taken with NIRCam, NIRSpec and MIRI. The NIRCam data had a resolving power of $R \simeq 1600$ and was only available for NIR38; NIRSpec provided a higher resolution of $R \simeq 2600$; and the data taken with MIRI had a poorer spectral resolution of $R \simeq 100$, since it was taken with the low resolution spectrograph (LRS) mode. We used the final reduced JWST spectra from the work by \cite{McClure2023}. Those final spectra can be reproduced by using the preprocessing modules built within AICE. The details of the reduction process can be found in Appendix \ref{appendix:pre-processing}.

Figure \ref{figure:aice-test} shows the resulting spectra for NIR38 and J110621 in the left panels (black curve) and the predictions of AICE for the ice composition in the right panel (grey bars). In the right panels of Fig. \ref{figure:aice-test} we also compare the AICE results of the ice fractional composition and temperature towards the two backgrounds stars with the results  derived by \cite{McClure2023} using \textsc{Eniigma} \citep{Rocha2021}, and \cite{Dartois2024} using radiative transfer calculations.  All these values are presented in Table \ref{table:aice-test}, which also presents the composition as ratios with respect to \ch{H2O}.  The predictions by AICE for the composition of the ices agree well with those obtained by \textsc{Eniigma} in \citet{McClure2023}. However, slightly larger discrepancies are found between our results and those reported by \cite{Dartois2024}. We note that the ice chemical composition given by \cite{Dartois2024} was initially assumed in their calculations.\footnote{\cite{Dartois2024} used a laboratory spectrum with an overall ice composition of H$_2$O$\,:\,$CO$_2$$\,:\,$CO$\,:\,$CH$_3$OH$\,:\,$NH$_3$$\,:\,$CH$_4$$\,:\,$OCN$^-$ with ratios 100:25:15:5:3:1:1, plus an additional pure CO contribution with a ratio with respect to the water fraction of the previous mixed ice of \ch{H2O}$\,:\,$CO 100:42, therefore having a total  \ch{H2O}$\,:\,$CO proportion of 100:57.}  As for the temperature of the ice, AICE tends to overestimate it for both sources. However, the ice temperature is a delicate variable,  as discussed in Section \ref{section:discussion}.

One last test consists of adding the AICE fractional compositions to check whether the total contribution lies close to 100 \%. The result is $88 \pm 5$ \% for both background stars. The total composition is close to 100 \%, but the difference suggests a slight underestimation of some of the ice molecular fractions by AICE. Additionally, it could indicate that the observed spectrum includes additional species different from the six molecules on which AICE has been trained.

As mentioned in Section \ref{section:training}, the uncertainties in the AICE results are estimated as the standard deviation of the predictions of ten different sets of neural networks (corresponding to different random seeds). We also need to take into account the uncertainties associated with the observed JWST spectra and hence, we propagate the observational uncertainties of the spectrum into the predictions of our model using the Python library RichValues (see Appendix \ref{appendix:richvalues} for more details). Then, we add this contribution quadratically to the standard deviation in the predictions for each variable. This observational contribution is overall negligible for NIR38 but it is significant for J110621, representing between 13\,\% and 33\,\% of the standard deviations of the neural networks depending on the variable.

\subsection{Comparison of the AICE predicted column densities}
\label{section:column-densities}

To obtain the predicted AICE column densities, we derive the column density of one of our targeted molecules and scale this value by using the fractional compositions given by AICE. The column density of one molecular species is calculated by integrating one of the molecule's IR bands in optical depth (or absorbance, in which case we introduce a factor of ln$\,$10) and dividing this value by the corresponding band strength. In our case, we chose the \ch{CO} band at 4.68 $\upmu$m, with a band strength of $1.4 \times 10^{-17}$ cm molecule$^{-1}$ at 14 K 
according to \citet{Bouilloud2015}. This band is selected because it is quite isolated, has a high signal-to-noise ratio, and corresponds to a species with a relatively high molecular fraction. To obtain the \ch{CO} column density, we integrate the observed band for both background stars, using the built-in AICE band integrator (see Appendix \ref{appendix:aice-integrator}). In this process, we exclude the central part of the band to account for possible saturation effects and subtract a local first-order baseline. The derived column densities can be found in Table \ref{table:aice-column-densities},  where they are compared to the values obtained by \cite{McClure2023} using the {\sc Eniigma} tool. The uncertainties were calculated by using Monte Carlo simulations with the Python library RichValues (Appendix \ref{appendix:richvalues}). The column densities obtained by AICE are all compatible with the estimates by {\sc Eniigma} within the uncertainties.

However, we stress that, as explained by \cite{Rocha2022}, integrating the bands and scaling by the band strength can involve uncertainties  up to $\sim$20 \% (or even more; e.g. \citealp{Bisschop2007, Oberg2007}). This is due to the fact that molecular band strengths of ice mixtures can be different from those of pure ices, and because they can also change with temperature. Indeed, differences of this order can be seen in the different band strengths reported by \cite{Bouilloud2015} for the same molecular species. 

\renewcommand*{\arraystretch}{1.5}
\begin{table}
\caption{Comparison of the ice column densities predicted by AICE and \textsc{Eniigma} \citep{McClure2023} for the background stars NIR38 and J110621.}
\label{table:aice-column-densities}
\centering
\resizebox{\hsize}{!}{
\begin{tabular}{lcccc}
\hline 
\multicolumn{1}{c}{} & \multicolumn{4}{c}{\textbf{column density ($10^{18}$ $\mathrm{cm}^{-2}$)}}\tabularnewline
\cline{2-5}
\multirow{2}{*}{\textbf{molecule}} & \multicolumn{2}{c}{\textbf{NIR38}} & \multicolumn{2}{c}{\textbf{J110621}}\tabularnewline
\cline{2-5}
 & \textbf{AICE} & \textbf{\textsc{Eniigma}} & \textbf{AICE} & \textbf{\textsc{Eniigma}}\tabularnewline
\hline 
{\bf $\mathbf{H_{2}O}$} & $7.4_{-1.3}^{+2.0}$ & $6.9_{-1.1}^{+1.9}$ & $10_{-2}^{+4}$ & $13.4_{-1.9}^{+1.3}$ \tabularnewline 
{\bf CO} & $2.0_{-0.4}^{+0.4}$ & $3.0_{-0.4}^{+0.6}$ & $2.9_{-0.7}^{+0.7}$ & $3.7_{-0.4}^{+0.6}$ \tabularnewline 
{\bf $\mathbf{CO_2}$} & $1.1_{-0.3}^{+0.4}$ & $1.38_{-0.20}^{+0.20}$ & $1.0_{-0.3}^{+0.4}$ & $1.74_{-0.22}^{+0.21}$ \tabularnewline 
{\bf $\mathbf{CH_{3}OH}$} & $0.56_{-0.25}^{+0.25}$ & $0.61_{-0.11}^{+0.11}$ & $0.6_{-0.3}^{+0.4}$ & $0.51_{-0.09}^{+0.19}$ \tabularnewline 
{\bf $\mathbf{NH_{3}}$} & $0.21_{-0.14}^{+0.16}$ & $0.30_{-0.03}^{+0.22}$ & $0.27_{-0.19}^{+0.24}$ & $0.66_{-0.06}^{+0.15}$ \tabularnewline 
{\bf $\mathbf{CH_{4}}$} & $0.30_{-0.11}^{+0.14}$ & $0.18_{-0.01}^{+0.02}$ & $0.37_{-0.18}^{+0.24}$ & $0.25_{-0.03}^{+0.01}$ \tabularnewline 
\hline 
\end{tabular}
}
\begin{justify}
\footnotesize{{\bf Note.} We used the column density of solid \ch{CO} to scale the fractional compositions predicted by AICE to derive the rest of molecular ice column densities.}
\end{justify}
\end{table}
\renewcommand*{\arraystretch}{1.0}

\subsection{Effects of band saturation}
\label{section:band-saturation}

As mentioned in Section \ref{section:column-densities}, some of the most intense molecular bands shown in Fig. \ref{figure:aice-test} may suffer from saturation effects, yielding a flat shape at the top of the IR bands in the observed spectrum. This is particularly critical for J110621 because it presents the highest extinction value ($A_{\rm V}$ = 95 mag) of the two stars studied by \citet[][]{McClure2023}. Therefore, in this section we evaluate whether our AICE results are affected by band saturation.  

A first look at the fractional ice compositions and ice column densities predicted by AICE (Tables \ref{table:aice-test} and \ref{table:aice-column-densities}) reveals that our predictions for \ch{H2O}, CO, and \ch{CO2} are overall consistent with previous calculations \citep{McClure2023,Dartois2024}. The \textsc{Eniigma} results indeed take into account the possible saturation of the bands. When looking at the spectra, however, the \ch{CO2} band at 4.27 $\upmu$m towards J110621 clearly shows saturation and hence, it seems that the AICE neural networks have learnt to recognise not only the relative intensities of the bands but also the full band profile shapes.

To better assess the sensitivity of our tool to band saturation, we retrained our neural networks with an augmented dataset in which we simulated the effect of band saturation. To do this, we used 3 copies of each spectrum in our training subset, applying a percentage of saturation (20\,\%, 40\,\% and 60\,\%) by cropping the absorbance values to those fractions of the maximum value of absorbance. We then retrained our models for our 7 targeted variables and our 10 random splits, obtaining a new set of model parameters (see Appendix \ref{appendix:saturation}). Interestingly, the validation errors are overall slightly lower than in the standard version of AICE.

The comparison between the AICE standard model and the AICE version with saturation effects is displayed in Table \ref{table:aice-saturation}. From this table, we find that both versions (AICE and AICE-sat) present consistent results within the uncertainties. However, we find some systematic differences between the predictions by both models. This new model provides larger predictions for CO and \ch{CH4,} while its predictions are lower for \ch{CH3OH} and \ch{NH3}. As for the predicted total sum, AICE-sat predicts $88 \pm 3$ \% for both NIR38 and J110621, which is similar than for the standard AICE model.

We note that the mentioned trends are observed for both NIR38 and J110621, even though the latter is expected to be more affected by saturation than NIR38, especially for the \ch{CO2} band at 4.27 $\upmu$m. In particular, the prediction by AICE-sat for the \ch{CO2} fraction remains very similar with respect to the one by the standard AICE, for both NIR38 and J110621, which suggests that the original version of AICE already learnt how to identify saturation (at least at these levels of uncertainty).  Comparing the predictions by AICE-sat with previous observations (see Table \ref{table:aice-test}), it seems that the new predictions are clearly worse for \ch{NH3}, \ch{CH4} and temperature. Hence, we  favour the use of the standard version of AICE over the version AICE-sat. Future versions of AICE could explore other ways to better treat IR band saturation to improve the performance of AICE both in the validation and in its application to real astronomical data.

\renewcommand*{\arraystretch}{1.2}
\begin{table}
\caption{Comparison between the predictions of the standard AICE model and the version trained with saturation effects (AICE-sat) towards NIR38 and J110621.}
\label{table:aice-saturation}
\centering
\resizebox{\hsize}{!}{
\begin{tabular}{lcccc}
\hline 
 & \multicolumn{2}{c}{\textbf{NIR38}} & \multicolumn{2}{c}{\textbf{J110621}}\tabularnewline
\cline{2-5}
{\textbf{prediction}} & {\textbf{AICE}} & {\textbf{AICE-sat}} & {\textbf{AICE}} & {\textbf{AICE-sat}}\tabularnewline
\hline 
{\bf $\mathbf{H_{2}O}$} (\%) & $56 \pm 3$ & $56.2 \pm 1.9$ & $59 \pm 3$ & $59 \pm 3$ \tabularnewline 
{\bf CO} (\%) & $15 \pm 3$ & $16.7 \pm 2.2$ & $16 \pm 4$ & $18.8 \pm 2.5$ \tabularnewline 
{\bf $\mathbf{CO_2}$} (\%) & $8.4 \pm 1.7$ & $8.1 \pm 0.9$ & $5.8 \pm 1.0$ & $5.9 \pm 0.6$ \tabularnewline 
{\bf $\mathbf{CH_{3}OH}$} (\%) & $4.2 \pm 1.7$ & $3.4 \pm 1.0$ & $3.4 \pm 1.5$ & $2.5 \pm 0.7$ \tabularnewline 
{\bf $\mathbf{NH_{3}}$} (\%) & $1.6 \pm 1.1$ & $0.9 \pm 0.6$ & $1.5 \pm 1.1$ & $1.0 \pm 0.7$ \tabularnewline 
{\bf $\mathbf{CH_{4}}$} (\%) & $2.3 \pm 0.8$ & $2.9 \pm 1.2$ & $2.1 \pm 1.0$ & $3.0 \pm 1.0$ \tabularnewline 
{\bf temp.} (K) & $23 \pm 4$ & $23 \pm 5$ & $19.0 \pm 2.0$ & $25 \pm 11$ \tabularnewline 
\hline 
\end{tabular}
}
\end{table}
\renewcommand*{\arraystretch}{1.0}

\subsection{Effects of changes in the spectral range}
\label{section:aice-lite}

Throughout this work, we use the same spectral range, from 4000 cm$^{-1}$ to 980 cm$^{-1}$ (2.5-10.2 $\upmu$m). In this section, we evaluate the performance of AICE when the observed spectrum presents a different wavelength coverage, especially if it is smaller. For this purpose, we retrained our neural networks using a smaller range, from 4000 cm$^{-1}$ to 2000 cm$^{-1}$ (2.5--5.0 $\upmu$m), which yields a total of 2001 points per spectrum and reduces the number of parameters on the first hidden layer of the networks.  In addition, we reduced the number of neurons on this first hidden layer from 120 to 90, given the reduced number of input points. After retraining our models, we found good results in the validation, although there was a slightly higher dispersion  and hence higher validation errors  (see Fig. \ref{figure:validation-lite}). 

The comparison between the standard version of AICE and this new version (AICE-lite), shown in Table \ref{table:aice-lite}, reveals that both sets of predictions are consistent within the uncertainties  for temperature and the main species \ch{H2O}, CO, and \ch{CO2}, as well as for \ch{CH3OH}. However, the predictions by AICE-lite are higher and more uncertain for the minor species \ch{CH4} and (especially) \ch{NH3}  compared to the predictions by the standard version of AICE (see Table \ref{table:aice-lite}). This can be due to the fact that both species do not have any strong feature in the range 4000-2000 cm$^{-1}$. Therefore, although AICE-lite retains a relatively good performance for abundant species such as H$_2$O, CO, and CO$_2$, training  AICE with a broader wavelength coverage provides more accurate results for \ch{CH3OH}, \ch{NH3} and \ch{CH4}.  Interestingly, the temperatures predicted by AICE-lite are slightly lower than the standard version of AICE, which could indicate that the spectra of these sources in the range of 4000--2000 cm$^{-1}$ (especially NIR38) is more compatible with lower temperatures compared with the part of 2000--980 cm$^{-1}$. Lastly, the predicted sums of the six targeted species are $91 \pm 5$ \% for NIR38 and $92 \pm 6$ \% for J110621, which is similar to those obtained with the AICE and AICE-sat models.

This shows that AICE can be successfully retrained to work with different spectral ranges. In fact, we could choose any range, providing that enough experimental data are available. In the case that the  spectra to be analysed has a slightly different spectral range than that of the training dataset, or even a different resolution, AICE can still work and predict the ice composition and temperature, but will present larger errors if the range and resolution of the input spectra significantly differ from those of the training dataset.

\renewcommand*{\arraystretch}{1.21}
\begin{table}
\caption{Comparison between the predictions of the standard AICE and the version trained in a smaller range of 4000-2000 cm$^{-1}$ (AICE-lite) towards NIR38 and J110621.}
\label{table:aice-lite}
\centering
\resizebox{\hsize}{!}{
\begin{tabular}{lcccc}
\hline 
 & \multicolumn{2}{c}{\textbf{NIR38}} & \multicolumn{2}{c}{\textbf{J110621}}\tabularnewline
\cline{2-5}
{\textbf{prediction}} & {\textbf{AICE}} & {\textbf{AICE-lite}} & {\textbf{AICE}} & {\textbf{AICE-lite}}\tabularnewline
\hline 
{\bf $\mathbf{H_{2}O}$} (\%) & $56 \pm 3$ & $53.2 \pm 2.1$ & $59 \pm 3$ & $54 \pm 3$ \tabularnewline 
{\bf CO} (\%) & $15 \pm 3$ & $12 \pm 3$ & $16 \pm 4$ & $16 \pm 4$ \tabularnewline 
{\bf $\mathbf{CO_2}$} (\%) & $8.4 \pm 1.7$ & $7.5 \pm 1.3$ & $5.8 \pm 1.0$ & $5.9 \pm 1.2$ \tabularnewline 
{\bf $\mathbf{CH_{3}OH}$} (\%) & $4.2 \pm 1.7$ & $4.3 \pm 2.5$ & $3.4 \pm 1.5$ & $5 \pm 3$ \tabularnewline 
{\bf $\mathbf{NH_{3}}$} (\%) & $1.6 \pm 1.1$ & $11.1 \pm 2.3$ & $1.5 \pm 1.1$ & $8 \pm 3$ \tabularnewline 
{\bf $\mathbf{CH_{4}}$} (\%) & $2.3 \pm 0.8$ & $2.8 \pm 1.2$ & $2.1 \pm 1.0$ & $3.2 \pm 1.5$ \tabularnewline 
{\bf temp.} (K) & $23 \pm 4$ & $17 \pm 4$ & $19.0 \pm 2.0$ & $21 \pm 7$ \tabularnewline 
\hline 
\end{tabular}
}
\end{table}
\renewcommand*{\arraystretch}{1.0}

\section{Discussion}
\label{section:discussion}

In this work, we demonstrate that artificial neural networks can be trained to identify and quantify the molecular composition of ices from IR absorption spectra. Our tool AICE gives accurate and fast predictions of the ice fractional composition when used to analyse laboratory spectra and also astronomical observations. The obtained results of the ice fractional composition towards the background stars NIR38 and J110621 in Chamaeleon I, observed with the JWST, are in agreement with previous estimates obtained using other methods (see Sections \ref{section:aice-astro-test} and \ref{section:column-densities}). However, previous estimates with {\sc Eniigma} use the band strengths of pure species to derive ice column densities in the observed astronomical mixtures. As discussed in Section 3.3, in the mixtures the band strengths could change up to $\sim$20\,\%. AICE has the advantage of implicitly considering band strength variations because the neural networks are trained with experimental ice mixtures, whose bands shapes, positions and intensities contain such information.  However, it is true that in order to create our augmented dataset with linear combinations from pure ices, we had to use molecular band strengths from pure ices.\footnote{ We note that we used the exact same band strengths as \cite{McClure2023} except for \ch{H2O} and CO, where we used the corrected values reported by \cite{Bouilloud2015}.} Still, the majority of the spectra are from direct experiments (2/3 of the training dataset), and with more experimental data available, the use of the augmented data could be avoided.  Therefore, the accuracy of the ice fractional composition predicted by the neural networks of AICE depends  mainly on the accuracy of the labeled composition of the experimental ice mixtures used for training. In addition, predictions of AICE do not seem to be sensitive to low or intermediate levels of band saturation, as explained in Section \ref{section:band-saturation}.

After all this work, we conclude that a plausible interpretation of the operations that AICE performs when analysing an input spectrum is as follows. In order to predict the fractional composition of a given species, it estimates how much absorption is being produced by it with respect to the total absorption observed in the spectrum, by taking into account the spectral shape of the spectral bands and their relative heights. In the case of the temperature prediction, it looks into the shape of the broader molecular bands, such as the ones from \ch{H2O}, \ch{CH3OH} and \ch{NH3} at $\sim$\,3.3 $\upmu$m and the \ch{H2O} bending band at $\sim$\,6.0 $\upmu$m. For obtaining the exact numeric predictions, AICE has learnt to scale the normalised absorbance values in the corresponding spectral bands (apparently avoiding their central parts, which are sensible to saturation), and those scaling relations were learnt from the training dataset. However, we should bear in mind that the exact actual behaviour of AICE may differ from this simplified explanation. Inferring its exact behaviour would be possible with techniques such as feature visualisation \citep{Olah2017} or just reverse engineering, but it would be quite time-consuming.

Several improvements could be implemented in AICE. First of all, more complex architectures could be further explored, especially one-dimensional convolutional neural networks. Secondly, and more important, the laboratory spectra dataset should be enlarged. Having a diverse and rich dataset  is essential for the accuracy of the predictions of the model. The list of target species could be enlarged to include complex organic molecules and molecular ions,  providing that enough experimental data exist. Current efforts focus on understanding whether RAIRS spectra (i.e. ice spectra obtained via reflection-absorption) could be used in large amounts to train future versions of AICE (e.g. spectra taken by \citealp{Kruczkiewicz2021}, or \citealp{Vitorino2024}). This would allow us to include the information from other databases such as the LIRA database,\footnote{\url{https://lerma.labo.cyu.fr/DR/index.php}} significantly enlarging the current AICE database. Additionally, enlarging the training dataset could enable the use of a single training-validation split (adding an extra division for testing), and therefore using a single neural network for each target variable, which would reduce the model size. Similarly, it could allow for the use of a single neural network instead of several ones for each target variable.

The comparison with other analysis methods \citep[][]{McClure2023,Dartois2024} shows that AICE is considerably faster than \textsc{Eniigma} and the radiative-transfer model of \cite{Dartois2022}. However, our model does not include any grain size parametrisation, as in the model of \cite{Dartois2022}, and it does not predict the ice composition for more complex molecules than CH$_3$OH, while \textsc{Eniigma} is capable of that. AICE could be trained with a larger dataset including a wider and more complete range of ice mixtures, and/or it could also include the effects produced by grain growth. Therefore, in principle, AICE could be expanded to include these features. Alternatively, our tool AICE could be implemented within other tools like the ones mentioned above, to provide a first guess of the composition of the ice, and then perform a more detailed calculation with \textsc{Eniigma} and \cite{Dartois2022}'s model, to make a precise fit of the observed astronomical spectrum. With respect to the spectra reduction, we note that the continuum determination is performed concatenating polynomials, so it might be prone to errors when obtaining the spectra in optical depth or absorbance. In addition, the removal of the silicate contribution is performed using a synthetic model, which might carry additional errors in the $\sim\,$8--10 $\upmu$m range, which is crucial to identify \ch{CH3OH}, \ch{NH3}, and more complex organics. For the comparison of the AICE results presented in Sections \ref{section:aice-astro-test} and \ref{section:column-densities}, we followed similar reduction steps to those performed by \cite{McClure2023}.

On the other hand, it is also important to mention that astronomical ices observed with the JWST present contributions from different dust grain populations along the line of sight, which  sum up and result in a  combined signal. This combined spectrum can differ from the type of spectra used to train AICE, since ices in laboratory are grown in a single sample with uniform temperature. We believe this is  one of the reasons behind the overestimation of the ice temperature by AICE for the ices measured towards NIR38 and J110621, since our model was trained with spectra that have a uniform temperature.  In fact, there are more considerations that should be taken into account about the temperature of ices.

The spectral signature of an ice (i.e. the shape and position of the bands) is modified by the  environment of the vibrators (the atoms), which is itself affected by the structure of the molecular film, namely, the way the molecules organise themselves in relation to each other. For example, for water deposited on a cold surface from the gaseous phase, the structure is very disordered, with a high surface-to-volume ratio, creating a complex topology with closed pores when the layer is large enough. As the temperature rises, annealing occurs, increasing the average number of coordination$\,$\footnote{The number of coordination in a molecule is the number of bounds to its neighbours, not only in a crystalline solid but in a general solid or cluster structure.} of each molecule and reducing the water-vacuum interfaces, which compacts the ice. The spectral signature also changes, as the mean number of coordination for each molecules increases (up to almost 4). However, this temperature-induced effect can be reproduced by other mechanisms. All that is needed to induce a change in the structure of the ice is to add energy, or to release energy. This energy can come from UV or even IR radiation \citep{Noble2020}, it can be deposited by energetic particles (electrons or ions) or simply released during a reaction (for example by H+H; \citealp{Accolla2011}). 

Therefore, the temperature value that is predicted by our tool AICE does not univocally represent the real temperature of the ice, but the temperature to which it has been brought/annealed. In the laboratory, we could obtain the same spectrum by going back down in temperature, and also without increasing the temperature but by depositing it in a different way (e.g. by jet; \citealp{Noble2024}), or by modifying it  by any energetic processing, even if slight. As our temperature variable does not really represent the temperature of the ice but rather the way in which it was annealed during the experiments, it is not surprising that our model suggests that the best description is an ice with the same signature as an ice annealed to a slightly higher temperature, which is simply due to the formation history of interstellar films, which is not the same as the one from a simple deposition at 10 K, but has been partly processed (upon its formation, chemical activity, radiative field  and energetic processing).

\section{Conclusions}

We have developed a machine learning model based on artificial neural networks that is able to predict the ice composition corresponding to an input IR absorption spectrum. The model has been packed into a user-friendly Python software called AICE and  trained to identify the temperature of the ice and its composition for the most commonly detected molecules (i.e. \ch{H2O}, \ch{CO}, \ch{CO2}, \ch{CH3OH}, \ch{NH3}, and \ch{CH4}) using hundreds of experimental ice absorption spectra. We computed the error of the model, obtaining average values of 2.5--2.7 \% for the fractional composition and 10.8--12.8 K for the temperature. When tested with astronomical observations taken by JWST, our model's predictions are in agreement with previous estimates obtained by other authors \citep{McClure2023, Dartois2024} within the uncertainties, with the exception of the temperature, which is slightly overpredicted. The computation time is very low (i.e. $\sim$\,0.1 s per input spectrum), which  allows us to automatically analyse hundreds of astronomical spectra measured with JWST and perform a statistically robust analysis. In addition, AICE is a versatile tool that could also be implemented or used in combination with other ice analysis tools such as \textsc{Eniigma}.

We performed tests using a dataset that reflects possible saturation effects in the bands of H$_2$O, CO, and CO$_2$, which demonstrated that AICE is little sensitive to these effects, since it learns to recognise the actual ice composition by focussing on the whole shape of the molecular band (including the band wings) and not on the relative peak intensity only. Additional tests reveal that AICE performance is optimal for the determination of the ice composition for species such as H$_2$O, CO, CO$_2$, and CH$_3$OH when using a reduced spectral range similar to the one provided by the JWST instruments NIRCam and NIRSpec.

Lastly, we also included an interactive toolkit within AICE which can be used by laboratory astrochemists to remove the experimental baselines and possible contaminations present in laboratory ice absorbance spectra, in an easy and semi-automated way. This could help in expanding the list of clean and reduced IR spectra available in public databases. Future versions of AICE are planned for upload\ to GitHub.

\section*{Data availability}

AICE is publicly available on GitHub to download.\footnote{\url{https://github.com/andresmegias/aice/}} In addition, a file of supplementary data is available in Zenodo,\footnote{\url{https://zenodo.org/records/16902313}} containing a table that lists all the experiments used in the training dataset of AICE.

\begin{acknowledgements}
We thank an anonymous referee for their useful comments and suggestions, which improved the quality of this paper. We thank M. E. Palumbo and A. Jim\'enez-Escobar for providing additional IR spectra from LASp (Italy) and NSRRC (Taiwan), respectively. We also thank C. del Burgo Olivares for his discussions about IR spectroscopy and F. Lesoil for testing the AICE Interactive Toolkit. In addition, we would like to aknowledge all the laboratory work done by the different authors of the experiments used in this paper. I.J-.S and A.M. acknowledge funding from grant PID2022-136814NB-I00 funded by the Spanish Ministry of Science, Innovation and Universities / State Agency of Research, MCIU/AEI/10.13039/501100011033 and by ERDF/EU. A.M. aknowledges support from grant PRE2019-091471 under project number PID2022-136814NB-I00, funded by MCIU/AEI/10.13039/501100011033, and by `ESF, Investing in your future'. B.M. acknowledges funding from grant PID2023-146415NB-I00, funded by the Spanish Ministry of Science, Innovation and Universities. This work is supported by ERC grant OPENS, GA No. 101125858, funded by the European Union. Views and opinions expressed are however those of the authors only and do not necessarily reflect those of the European Union or the European Research Council Executive Agency. Neither the European Union nor the granting authority can be held responsible for them. This work has received suport from Cergy Paris Universit\'e through the AstroCY program, and from the French Agence Nationale de la Recherche (ANR) through the SIRC project (grant ANR-SPV2024482020-2024). This project has received funding from the European Research Council (ERC) under the European Union’s Horizon 2020 research and innovation programme (grant agreement No. 291141 MOLDISK).
\end{acknowledgements}

\bibliography{refs}
\bibliographystyle{aa}

\begin{appendix}

\section{Reduction of astronomical ice absorption spectra}
\label{appendix:pre-processing}

\begin{figure*}
\centering
\includegraphics[trim={0 10 0 24}, clip, width=0.8\hsize]{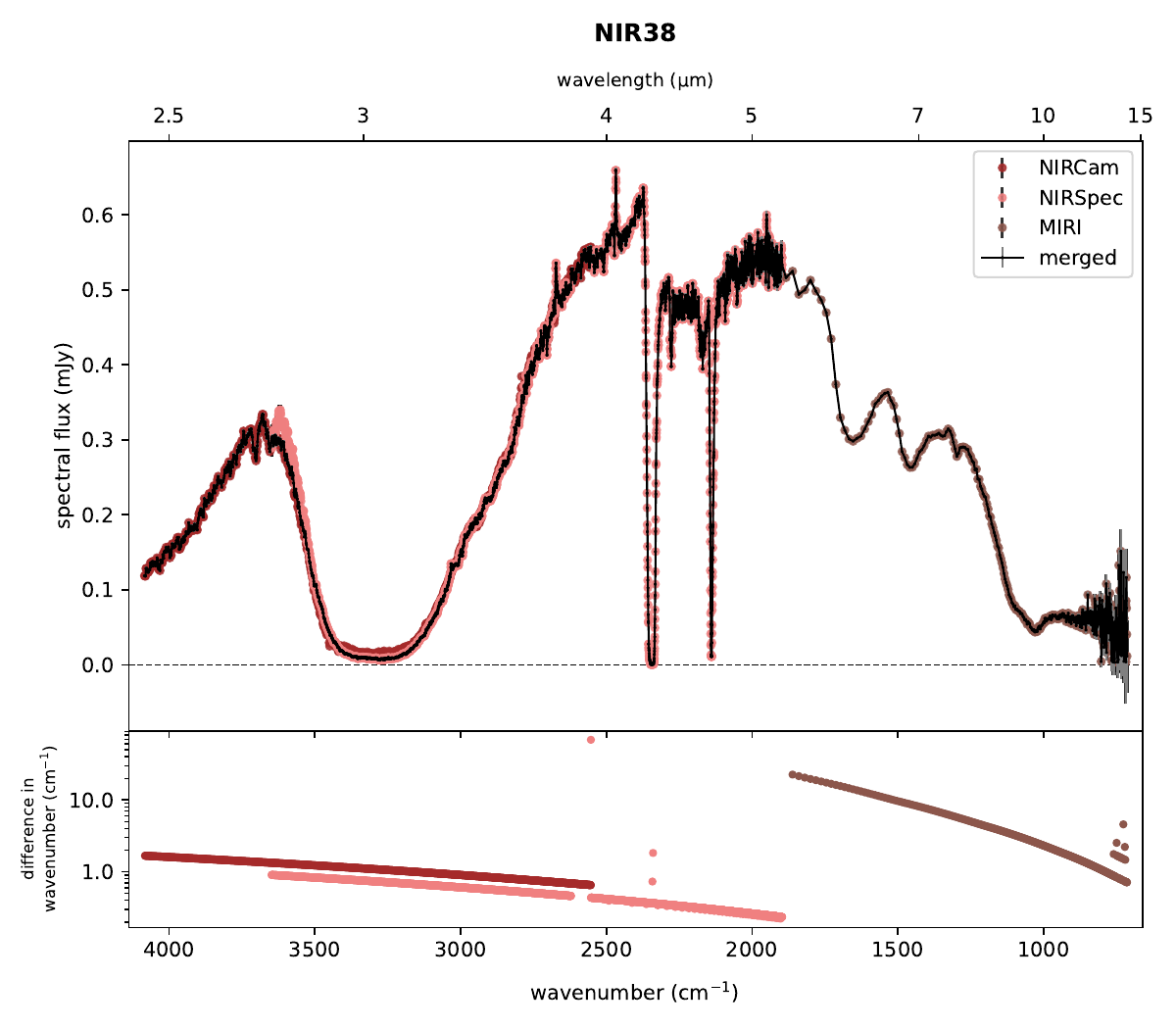}
\caption{Spectra taken by JWST \citep{McClure2023} towards the background star NIR38 in the molecular cloud Cha I, with its three instruments, and resulting combined spectrum. Lower panel shows the spacing in wavenumber between each point of the spectra, to visualise the resolution of the data. The points with larger wavenumber differences correspond to gaps in the original spectra.}
\label{figure:aice-pre1-merging}
\end{figure*}

\begin{figure}
\centering
\includegraphics[trim={0 230 0 20}, clip, width=\hsize]{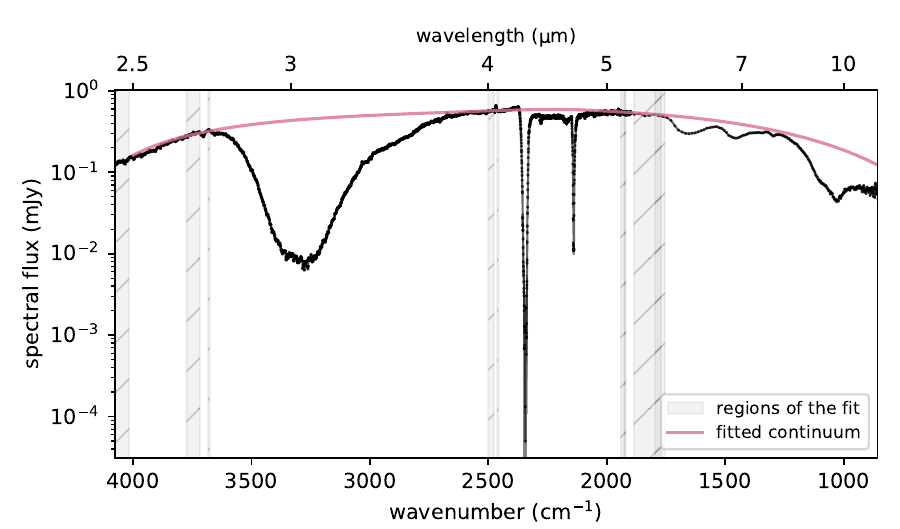}
\caption{Spectrum taken by JWST towards the background star NIR38 in the molecular cloud Cha I, and fitted continuum, used to compute the absorbance spectrum. The fit was performed by \cite{McClure2023} joining several polynomials fitted within the dashed regions and extrapolating to the rest.}
\label{figure:aice-pre2-continuum}
\end{figure}

\begin{figure}
\centering
\includegraphics[trim={0 230 0 20}, clip, width=0.98\hsize]{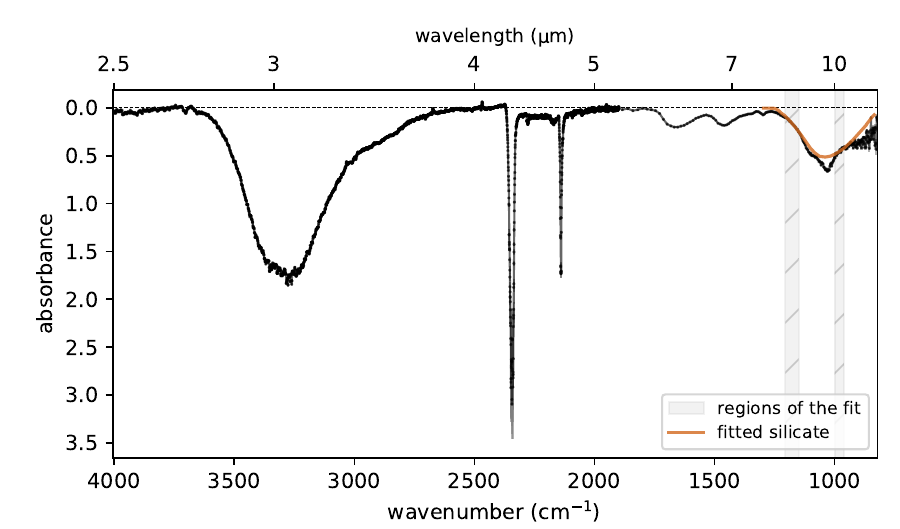}
\caption{Spectrum in absorbance derived from data by JWST towards the background star NIR38 in the molecular cloud Cha I, and fitted silicate contribution, which has to be subtracted to obtain the ice spectrum. The fit was performed by \cite{McClure2023} using the software OpTool \citep{Dominik2021} in the dashed regions.}
\label{figure:aice-pre3-silicates}
\end{figure}

Astronomical observations of ices consist of the spectral flux of light received towards a background source, $\mathcal{F}$. To convert it to optical depth ($\tau$) or absorbance ($A$), we need to know or estimate the non-absorbed continuum, $\mathcal{F}_0$, to then apply the logarithm of the ratio of fluxes:
\begin{equation}
    \indent \indent
    \tau \; = \; \ln{(\mathcal{F}_0 / \mathcal{F}}) \; = \; A \ln(10)   \;.
\end{equation}
The continuum can be estimated by different means, although a common practice is to fit a black-body curve or a polynomial curve (e.g.: \citealp{Rocha2021}; \citealp{McClure2023}). In this scale, the contribution of several macroscopic layers of ice combines additively. This allows us to derive the quantity of ice (column density, $N$) integrating the area of a certain band and scaling by the experimentally derived area (band strength, $S$):
\begin{equation}
    \indent \indent
    N \; = \; \frac{\int_{\tilde \nu_1}^{\tilde \nu_2} \tau(\tilde \nu) \, {\rm d} \tilde \nu }{S} \; = \; \frac{\ln(10)}{S} \int_{\tilde \nu_1}^{\tilde \nu_2} A(\tilde \nu) \, {\rm d} \tilde \nu \;,
\end{equation}
where $\tilde \nu$ is the wavenumber, that is, the inverse of the wavelength ($\lambda$), $\tilde \nu$ = 1/$\lambda$, and ($\tilde \nu_1$, $\tilde \nu_2$) is the extension of the band. Before searching for molecular features, one more step is needed: subtraction of the absorption due to the silicate-rich core of the dust grains, which has a noticeable effect in the $\sim\,$8--12 $\upmu$m range; this can be done with polynomials or with silicate models, like the ones provided in the software OpTool \citep{Dominik2021}.

These steps can be performed within AICE using its three preprocessing modules,  as we showcase in the next figures for the spectra of NIR38 and J110621 taken by JWST by \cite{McClure2023}. The first module allows us to merge spectra coming from different instruments to a single spectrum expressed in wavenumber (Fig. \ref{figure:aice-pre1-merging}). The second module allows us to fit several polynomials to the observed spectrum, in logarithmic scale and in different wavelength regions, to construct an estimation of the continuum, in order to convert the spectra from flux to absorbance (Fig. \ref{figure:aice-pre2-continuum}).  Instead of fitting the polynomials to the data, the module also allows us to load a precalculated continuum. The propagation of the uncertainties from flux to absorbance is done with the Python library RichValues (see Appendix \ref{appendix:richvalues}). Lastly, the third module of preprocessing allows us to fit a silicate contribution to the absorbance spectrum in certain regions with no expected ice bands (Fig. \ref{figure:aice-pre3-silicates}), using the software OpTool \citep{Dominik2021}.  It produces a synthetic spectrum and scales and shifts it to fit the observed spectrum in the fitted regions, using a  certain grain size and composition.  In the case of NIR38 and J110621, the grain size is 1 $\upmu$m and its composition is a combination of pyroxene and olivine (for more details, see the work by \citealp{McClure2023}). Again, instead of fitting the silicate contribution, this module also allows us to load a precalculated curve. 

Additionally, we should note that before applying the AICE model to a spectrum, it has to be resampled to a spectral range of 4000--980 cm$^{-1}$ with a resolution of 1 cm$^{-1}$. The rebinning of the spectrum is done with the help of the SpecRes Python tool by \cite{Carnall2017}. We supersample the part of the spectrum obtained with the MIRI instrument, which has a poorer resolution (LRS mode), to get a resolution of 1 cm$^{-1}$, artificially increasing the uncertainties to make them consistent with the original data.

\section{Basics of artificial neural networks}
\label{appendix:neural-networks}

This section explains in detail the definition of artificial neural networks, focussing on multi-layer perceptrons, and how the training procedure works.

\subsection*{Mathematical definition}

A multi-layer perceptron is a mathematical function based on the concatenation of linear combinations and non-linear activation functions \citep{Murtagh1991}. Starting from an input vector of size $n$, this function $f$ applies a set of operations to them in several steps or layers, ending with an output vector, $\boldsymbol{\hat y} = f(\boldsymbol{x})$. In the first layer, the inputs are converted to an intermediate vector, $\boldsymbol{a}$. Each one of its elements is the result of a different linear combination of the input parameters, plus a non-linear function $\phi$ applied to it \citep{Grosan2011}:
\begin{equation}
    \indent \indent
    a_j \;=\; \phi \left ( \, \sum_{i=1}^{n} \, w_{ji} \; x_i \,+\, b_{j} \right ) \;,
\end{equation}
where $w_{ji}$ are the weights of the linear combination and $b_j$ is an offset, referred to as the bias. The activation function $\phi$ can adopt several forms, but a common use is the rectified linear unit (ReLU), defined as $\phi(x) = \max(0, x)$ \citep{Fukushima1975, Nair2010}.

Each one of the  mentioned operations constitute the neurons and $a_j$ are the activations of each neuron; all of them constitute the first hidden layer of the neural network. Then, the same operation (equation B.1) is applied to the activations of the first hidden layer as input values, but with different values of $w_{ji}$ and $b_{j}$, obtaining as a result a second hidden layer. Therefore, the activations of the layer, $l$, can be generalised as
\begin{equation}
    \indent \indent
    a_{lj} \;=\; \phi \left ( \, \sum_{i=1}^{n_{\,l-1}} \, w_{lji} \; a_{l-1,i} \,+\, b_{lj} \right ) \;,
\end{equation}
where $n_{\,l-1}$ is the number of activations of the previous layer and with  $a_{0,i} = \,x_i$. After a certain number $L$ of steps or layers, we end up with a final vector, $\boldsymbol{\hat y}$, which constitutes the final output of the function $f$, the output layer. In this last set of operations, where $a_{L,j} = \,\hat y_j$, the activation function can adopt different shapes depending on the desired format for the output. For example, if the outputs have to be between 0 and 1, a  sigmoid function can be used: $\phi(x) = 1 \,/\, (1 + {\rm e}^{-x})$; in another case, if the outputs have to be positive, we can still use a ReLU as activation function.

\subsection*{Training of a neural network}

An artificial neural network is a function, $f$, with several parameters: the weights, $w_{lji}$, and the biases $b_{lj}$, where the indices refer to the neuron $j$ of the layer $l$ and the previous activation $i$. We can see that even a small model with a few layers and neurons would have plenty of free parameters. 

Thanks to its multiple parameters and to the non-linear activation functions, artificial neural networks have a huge versatility to perform several different operations and transformations to the input data. In particular, they can be `trained' to imitate and learn a particular transformation or operation as long as we have several examples of it \citep{Larochelle2009}. For the sake of simplicity, let us assume that the output of the neural network is a single value, $\hat y$, and that the input is a set of values that we group into a vector $\boldsymbol{x}$, so that $\hat y = f(\boldsymbol{x})$. We start from a group of $m$ inputs, $\{\boldsymbol{x}_k\},$ and we also know the corresponding $m$ values after the transformation, $\{y_k\}$. The set of pairs ($\boldsymbol{x}, y$) is called training dataset. We want the neural network $f$ to return each value $y_k$ when applied to the corresponding vector $\boldsymbol{x}_k$. This translates into an optimisation problem, where we want to minimise the difference between the predictions of the model, $\hat y_k = f(\boldsymbol{x}_k)$, and the reference values, $y_k$. To quantify this difference, we define an error function, or loss function ($\mathcal{L}$), which can be the known mean squared error (MSE): 
\begin{equation}
\label{equation:loss}
    \indent \indent
    \mathcal{L} \; = \; \frac{1}{m} \, \sum_{k=1}^{m} \, ( \hat y_k - y_k)^2  \;,
\end{equation}
where the sum ranges over the examples in our training dataset. The process of optimising the parameters of $f$ (the weights and biases) to minimise the loss $\mathcal{L}$ is called training. This is analogous to a linear fit to a set of $m$ points ($x, y$), where the size of both the input and the output is 1. In case that both the input and the output are vectors, ($\boldsymbol{x}, \boldsymbol{y}$), we would use equation \ref{equation:loss} in a vectorial form, calculating the squared norm of the vectorial difference between $\boldsymbol{\hat y}_k$ and $\boldsymbol{y}_k$, that is, $|| \boldsymbol{\hat y}_k - \boldsymbol{y}_k ||^2$. In the training of a neural network, its parameters are initialised randomly and then modified in consecutive steps or \textit{epochs}, to explore the parameter space and find values that minimise the loss function when applying $f$ to the input vectors, $\boldsymbol{x}_k$. This process of finding the minimum of the loss function is known as \textit{gradient descent}, and in neural networks this is possible thanks to the \textit{backpropagation} algorithm \citep{Rumelhart1986, LeCun1998}. Gradient descent comprises several variations of the same method with the same objective. In all of them, there is a parameter called {\it learning rate} that controls how large the changes in the parameters of the network are after each training epoch.

Lastly, there is an additional step that must be done before starting the training. In order to avoid overfitting, the neural network should not be trained using the entire training dataset. Instead, we randomly select a small fraction of it (e.g. $\sim$10 \%) and separate it from the rest; this subset is called validation dataset, and will be used to quantify the progress in the training of our model. While the fitting will be performed on the training subset, the loss function will also be evaluated on the validation subset, informing about the performance of the model $f$ in new or unseen data. The training steps will be repeated in several epochs that will reduce the loss function in both subsets, on average, until the validation loss stops improving significantly.

\subsection*{Additional improvements}

We list here some additional procedures that can improve the training of neural networks, and that we used to train AICE:
\vspace{-0.15cm}
\begin{itemize}[leftmargin=12pt]
    \item {\it Split the training subset in mini-batches.} The loss function is evaluated for each mini-batch (a group of data elements), applying also the gradient descent to update the network parameters. In this way, the whole dataset is covered in several steps, which completes an epoch. Therefore, the training is accelerated, since the number of times the optimisation is performed is increased.
    \item {\it Use batch normalisation.} This means that, for each mini-batch, the output of a layer (activations, $a$) is standarised, by subtracting its mean along the batch ($\mu_a$) and dividing by its standard deviation ($\sigma_a$), and then a new linear transformation with two learnable parameters is applied \citep{Ioffe2015}. The transformed activation would be
    \begin{equation}
        \indent \indent
        \;\;\;\; a' \; = \; \gamma_a \, \frac{a  -  \mu_a}{\sqrt{\sigma_a^2 + \epsilon}} \, + \, \beta_a  \;, 
    \end{equation}
    where $\gamma_a$ and $\beta_a$ are the new trainable parameters (a weight and a bias) and $\epsilon$ is a small term introduced to avoid division by 0 (in our case, $\epsilon = 10^{-3}$). For inference (after the training), the mean and variance in the activations for the whole training dataset are used as $\mu_a$ and $\sigma_a^2$, since the split in batches is only done during training. This is applied for all the activations of the layer, therefore having a different set of ($\mu_a$, $\sigma_a$, $\gamma_a$, $\beta_a$) parameters for each neuron in the layer (for each activation $a$).
    \item  {\it Apply dropout to some activations.} This consists on setting a random fraction of the activations of a certain layer to 0, with a certain probability, which can help to reduce overfitting \citep{Srivastava2014}.
    \item {\it Reduce the learning rate along the training.} The learning rate defines the size of the changes in the parameters along the training of the networks. Reducing this value after a certain number of epochs with no reduction of the loss can help to improve the optimisation of the parameters. In fact, the learning rate in Adam is adaptive. The value selected by the user corresponds to the maximum learning rate, but it can be reduced along the optimisation, as we did in the training of AICE.
\end{itemize}

\section{Baseline removal for laboratory spectra}
\label{appendix:aice-toolkit}

To remove the experimental baselines from our laboratory training dataset, and for some additional processing, we used a tool included within AICE,  called AICE Interactive Toolkit. This software is a Python 3 script that allows us to read absorbance spectra in plain text files, plotting the input spectrum (or spectra) and allowing the user to interact via the mouse and keyboard to perform different actions. In the following, we present an example of the use of this tool to process the experiment \#12 from LIDA, corresponding to a mixture of \ch{H2O}$\,$:$\,$HCOOH (91:9), published originally by \cite{Bisschop2007}.

\begin{figure}
\centering
\includegraphics[trim={0 12 0 24}, clip, width=\hsize]{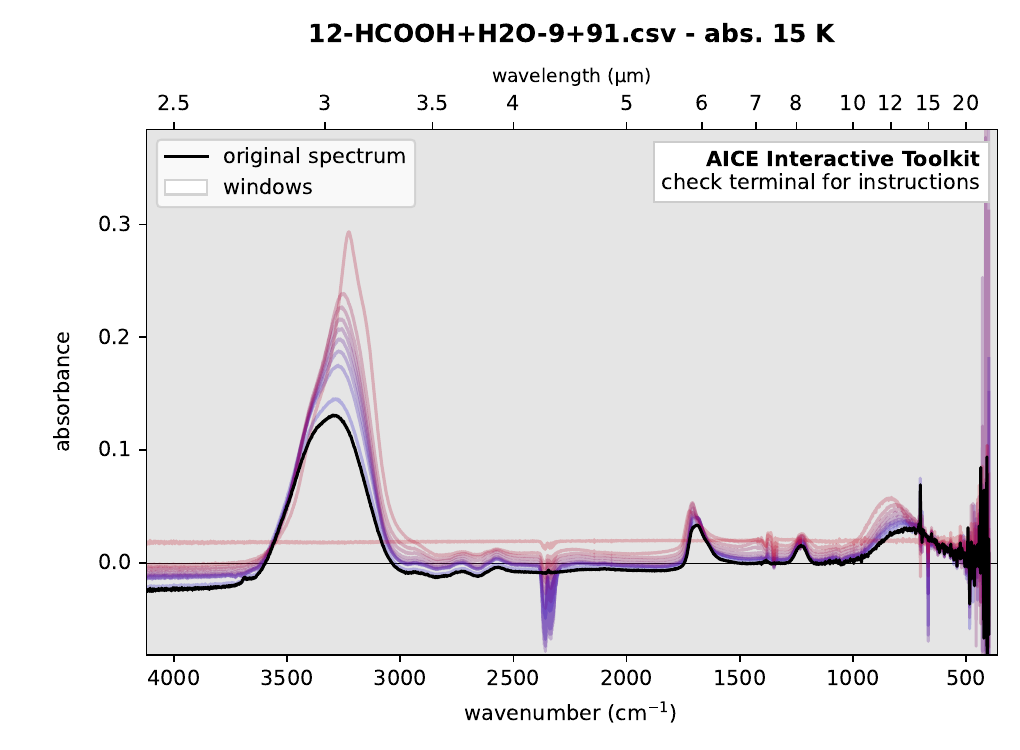}
\caption{Spectra obtained in the experiment \#12 of LIDA, from a mixture of \ch{H2O}$\,:\,$HCOOH (91:9), as seen with our AICE Interactive Toolkit. The curve in black is the spectrum at 15 K, and the rest of coloured curves correspond to spectra with increasing temperature, with a step of 15 K until 165 K (which is almost flat, as both species have desorbed). The spectra were originally published by \cite{Bisschop2007}.}
\label{figure:aice-toolkit-1}
\end{figure}

Figure \ref{figure:aice-toolkit-1} shows the spectra as seen  after launching our interactive tool from the terminal and loading the input data. We can see that each spectrum has a different offset or baseline, that should be removed. Beforehand, we have to exclude the data corresponding to more than 100 K since, as we comment in Section \ref{section:origin-data}, we focus on spectra with temperatures of $\leq$100 K. Additionally, we can spot some contamination and noise between 2500 and 2000 cm$^{-1}$ and at the very right of the spectra, which will be corrected after the baseline subtraction.
\begin{figure}
\centering
\includegraphics[trim={0 12 0 24}, clip, width=\hsize]{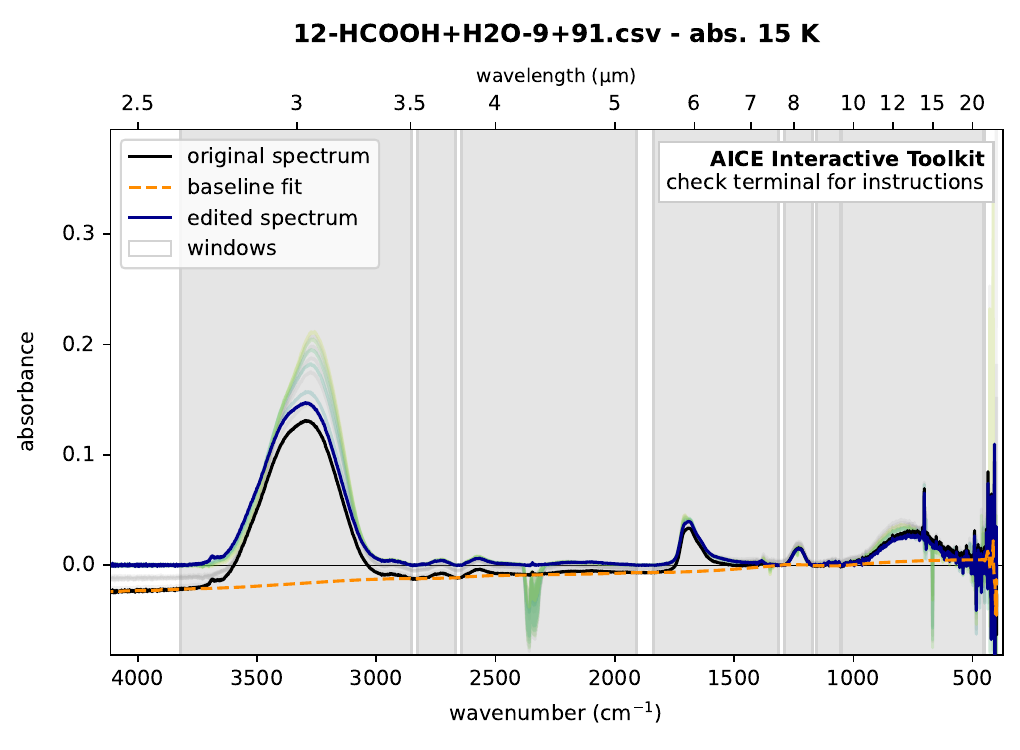}
\caption{Same as Fig. \ref{figure:aice-toolkit-1}, after removal of the spectra with temperatures greater than 100 K and after baseline subtraction. Now the curve in orange shows the fitted baseline (in the white spectral windows) for the spectrum at 15 K (black curve), and the blue curve shows the corresponding reduced spectrum. The rest of coloured curves are the reduced spectra for the rest of temperatures.}
\label{figure:aice-toolkit-2}
\end{figure}

\begin{figure}
\centering
\includegraphics[trim={0 12 0 24}, clip, width=\hsize]{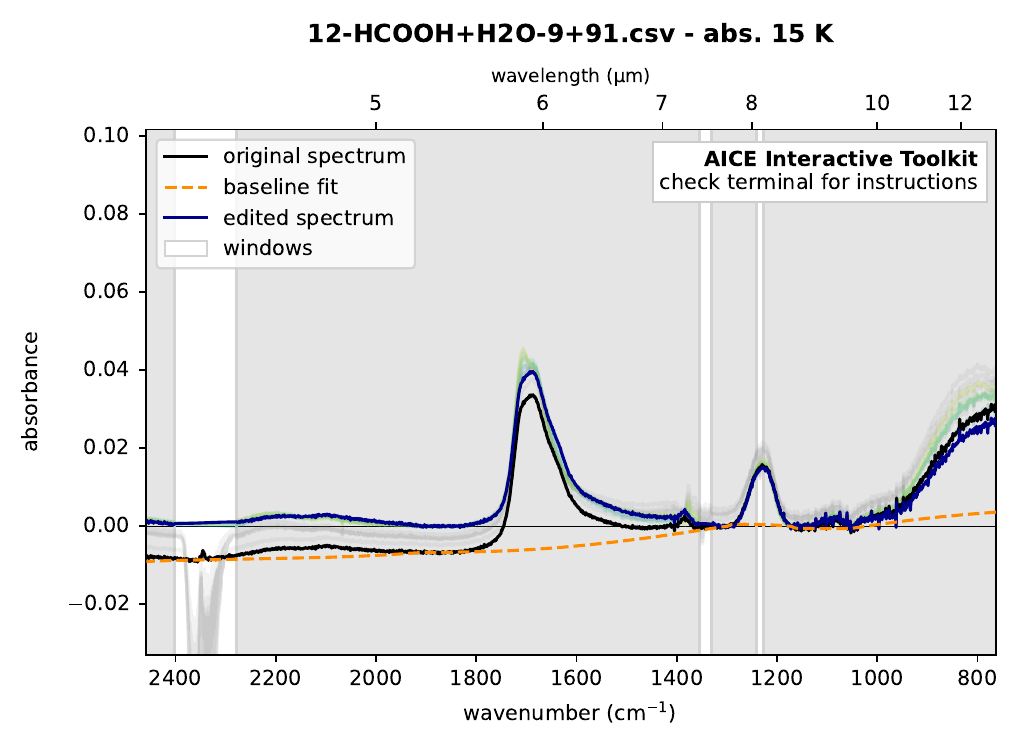}
\caption{Same as Fig. \ref{figure:aice-toolkit-2} in a zoomed region, after a smoothing the spectra with a size of 7 and interpolating two contaminated regions. The curve in black shows the original spectrum at 15 K, the curve in blue shows the reduced spectrum also at 15 K, and the dashed curve in orange is the fitted baseline.}
\label{figure:aice-toolkit-3}
\end{figure}

In order to fit the baseline for all the spectra in this experiment, we first select several spectral windows where the spectra should be flat, namely, with no absorption. After manually selecting the windows (with the cursor), we can now fit a baseline in those windows (for each recorded temperature) and reduce the spectra removing the fitted curves. The mathematical form of the fitted Gaussian is a spline, that is, a concatenation of polynomials, of third order. We use the function/class \textsf{UnivariateSpline} from Python's package SciPy, which requires as an input a smooth factor. In order to compute it, we require a different smooth parameter, $s$, that can be input by the user (by default, this parameter is $s$ = 25). We first smooth our spectra in the selected windows applying a rolling median with a size of $s$. Then, we compute the sum of the squared errors between the smoothed curve and the original spectrum, in the selected windows; this is the smooth factor that we input to \textsf{UnivariateSpline}. In this way, the condition to be satisfied to select the number of polynomials for the spline (the number of {\it knots}) is that the sum of the squared errors between the spline and the original spectrum (in the selected windows) is less than or equal to the sum of the squared errors previously calculated (between the spectrum and its smoothed version). Equivalently, we could use the mean squared error (MSE) instead of the sum of the squared errors. Figure \ref{figure:aice-toolkit-2} shows the result obtained after performing these steps, with the default value of $s$ = 25. We can now see that the reduced spectra have an absorption of zero in the selected regions. At the rightmost part of the spectrum, the baseline presents too much variation, but we do not try to fix this since we are interested in the 4000 to 980 cm$^{-1}$ range.

We have performed the most important steps to reduce the spectra from this experiment. However, additional corrections should be applied to the data. Again in the rightmost part of the spectra, too much noise can be seen. Since we do not have very thin absorption features in the spectra, we can apply a rolling mean to smooth the data in all the spectral range; otherwise, we should apply this excluding the regions with thin features. Therefore, we applied a smoothing with a size of 7 channels. After that, we focus on two regions where signs of contamination can be seen. The most noticeable one appears at around 2400--2300 cm$^{-1}$, centered in around 2340--2330 cm$^{-1}$. This is certainly very similar to the main absorption band of \ch{CO2}, so it seems that it is due to some gaseous \ch{CO2} present in the experimental chamber. Similarly, we see a pattern around 1400--1300 cm$^{-1}$. The left half of it (1400--1350 cm$^{-1}$) corresponds to an absorption band of HCOOH, but the second half (1350--1300 cm$^{-1}$) seems to be an artifact or a contamination whose origin we cannot identify. Additionally, there is an artifact at $\sim$1230 cm$^{-1}$. In order to remove all of these features, we select the three spectral windows where these contaminations/artifacts are present. Then, the tool removes the data in the selected regions and interpolates the spectra in the window edges using polynomials. Fig. \ref{figure:aice-toolkit-3} shows the resulting spectra, zoomed into the region of interest.

Once this is done, the spectra processing is virtually done. Optionally, we could artificially add random noise in the last spectral windows to avoid having a fully smooth curve in the interpolated regions.

\section{The augmented dataset for AICE}
\label{appendix:coefficients}
The training dataset used for AICE consists of 574 experimental spectra plus 282 synthetic spectra. This last part was constructed using pure ice spectra from LIDA, OCdb and Univap, that were normalised by its integrated column density and then combined linearly to form the 282 spectra.

\begin{table*}
\caption{Spectra and band strengths used to construct the augmented dataset of linear combinations of pure ice spectra.}
\label{table:pure-ices}
\centering
\resizebox{\textwidth}{!}{
\begin{tabular}{lccccr}
\hline 
\multirow{2}{*}{\textbf{Molecule}} & \multirow{2}{*}{\textbf{Experiment}} & \multirow{2}{*}{\textbf{Reference}} & \textbf{Integrated band} & \textbf{Band strength} & \multirow{2}{*}{\textbf{Reference}}\tabularnewline
 &  &  & (cm$^{-1}$) & (cm molecule$^{-1}$) & \tabularnewline
\hline 
\multirow{2}{*}{$\mathrm{H_{2}O}$} & \multirow{2}{*}{OCdb \#74--77 (10--100 K)} & \multirow{2}{*}{\cite{Hudgins1993}} & \multirow{2}{*}{$3330\pm400$} & \multirow{2}{*}{$2.2\cdot10^{-16}$} & \multirow{2}{*}{\cite{Gerakines1995}}\tabularnewline
 &  &  &  &  & \tabularnewline
\multirow{2}{*}{$\mathrm{CO}$} & LIDA \#75 (15--25 K) & \cite{VanBroekhuizen2006} & \multirow{2}{*}{$2139\pm5$} & \multirow{2}{*}{$1.4\cdot10^{-17}$} & \multirow{2}{*}{\cite{Gerakines1995}}\tabularnewline
 & Univap \#G1 (10 K) & \cite{Rocha2014} &  &  & \tabularnewline
\multirow{2}{*}{$\mathrm{CO_{2}}$} & \multirow{2}{*}{OCdb \#45--48 (10--70 K)} & \multirow{2}{*}{\cite{Hudgins1993}} & \multirow{2}{*}{$2343\pm15$} & \multirow{2}{*}{$1.1\cdot10^{-16}$} & \multirow{2}{*}{\cite{Gerakines1995}}\tabularnewline
 &  &  &  &  & \tabularnewline
\multirow{2}{*}{$\mathrm{CH_{3}OH}$} & \multirow{2}{*}{OCdb \#37--40 (10--100 K)} & \multirow{2}{*}{\cite{Hudgins1993}} & \multirow{2}{*}{$1027\pm50$} & \multirow{2}{*}{$1.8\cdot10^{-17}$} & \multirow{2}{*}{\cite{Hudgins1993}}\tabularnewline
 &  &  &  &  & \tabularnewline
\multirow{2}{*}{$\mathrm{NH_{3}}$} & \multirow{2}{*}{LIDA \#116 (10--70 K)} & \multirow{2}{*}{\cite{Taban2003}} & \multirow{2}{*}{$1070\pm60$} & \multirow{2}{*}{$2.1\cdot10^{-17}$} & \multirow{2}{*}{\cite{Sandford1993}}\tabularnewline
 &  &  &  &  & \tabularnewline
\multirow{2}{*}{$\mathrm{CH_{4}}$} & \multirow{2}{*}{OCdb \#42-44 (10--30 K)} & \multirow{2}{*}{\cite{Hudgins1993}} & \multirow{2}{*}{$1303\pm10$} & \multirow{2}{*}{$8.4\cdot10^{-18}$} & \multirow{2}{*}{\cite{Boogert1997}}\tabularnewline
 &  &  &  &  & \tabularnewline
\hline 
\end{tabular}
}
\begin{justify}
\footnotesize{{\bf Note.} The position of the integrated band is given as the centre plus/minus an approximate width. Band strengths were derived from the bibliographic work by \cite{Bouilloud2015}, taking the corrected values corresponding to the closest temperature to the experimental spectrum for each case. Note: for CO we joined the spectrum \#G1 from Univap with the spectra \#75 from LIDA, to increase the effective temperature range.}
\end{justify}
\end{table*}

Table \ref{table:pure-ices} shows information about the experiments corresponding these pure ice spectra and the integrated band used to derive the corresponding column density. All of these experiments provided spectra of pure ices for several temperatures, ranging from 10 to 100 K at maximum. Once the ice column density is derived for the spectra with lowest temperature, we divide each spectra in absorbance by the column density, obtaining the absorption cross-section (cm$^2$).

Now, we can create a linear combination of our six pure ice cross-sections, $\sigma_{m,T}(\tilde \nu)$ (with $m$ \,=\, \ch{H2O}, \ch{CO}, \ch{CO2}, \ch{CH3OH}, \ch{CH4}, \ch{NH3}), to construct  a synthetic spectrum, in terms of the wavenumber, $\tilde \nu$. Since we will further normalise the resulting spectra by their mean absorbance value, we will not care about the absolute values of the coefficients of the linear combinations, $\{c_m\}$, so we will define them between 0 and 1 and normalised so that their sum is equal to 1, for the sake of simplicity. We want to obtain a curve $y_{\boldsymbol{c},T}(\tilde \nu)$ corresponding to the spectrum with temperature $T$ and coefficients $\boldsymbol{c}$ = ($c_{\ch{H2O}}$, $c_{\ch{CO}}$, $c_{\ch{CO2}}$, $c_{\ch{CH3OH}}$, $c_{\ch{NH3}}$, $c_{\ch{CH4}}$). Therefore, the resulting spectrum will be:
\begin{equation}
    \indent \indent
    y_{\boldsymbol{c},T}(\tilde \nu) \; = \; \sum_m \, c_m \, \sigma_{m,T}(\tilde \nu) \;. 
\end{equation}
We note, however, that the temperatures $T$ at which the experimental spectra of the pure ices were recorded are not always the same. Therefore, we should find a way to make an interpolation to obtain the pure ice spectra at the desired temperature. For this, we create a linear interpolation along the variable $T$ of the available experimental cross-sections, for each targeted molecule, and inside the temperature range for each pure ice. For example, for \ch{H2O} we have curves for 10 K, 40 K, 80 K, and 100 K. If $T$ = 20 K, we will interpolate the curves at 10 K and 40 K, obtaining a result close to the cross-section at 10 K.

Once we have all our interpolated cross-sections in the dimension of temperature for all our six molecules, we can do a linear combination of them to obtain a synthetic mixture. However, choosing the possible values of the coefficients is not trivial.\footnote{A simple approach would be to generate 6 random numbers from a uniform distribution between 0 and 1, and divide them by the sum of all of them. However, in this way the probability of obtaining a normalised coefficient greater than $\sim$0.5 is very low.} First of all, the sum of the six coefficients must be always 1, $\sum_m c_m  =  1$. Then, we are interested in representing all the possible combinations of our six molecules in a balanced way, and exploring all the parameter space, that is, having a significant number of values of the coefficients between the whole range from 0 to 1. To do so, we will use the following algorithm:
\vspace{-0.2cm}
\begin{itemize}[leftmargin=12pt]
    \item We start with the molecule $m_1$. We will set its coefficient $c_1$ from a uniform distribution between 0 and 1.
    \item Then, we go for the molecule $m_2$. We will draw its coefficient $c_2$ from a uniform distribution between 0 and $1 - c_1$.
    \item We repeat the previous step until the second to last molecule, $m_5$. The coefficient will be drawn from a uniform distribution between 0 and $1 - c_1 - c_2 - c_3 - c_4$.
    \item The coefficient for the last molecule $m_6$ will be directly $c_6 = 1 - c_1 - c_2 - c_3 - c_4 - c_5$.
\end{itemize}
\vspace{-0.2cm}
In this way, we have obtained a combination of coefficients ($c_1$, $c_2$, $c_3$, $c_4$, $c_5$, $c_6$). To obtain a different combination of parameters, we just start again the algorithm, obtaining a new initial coefficient $c_1$ and, in each step, the rest of the coefficients.

With the previous algorithm, we will obtain a set of coefficients, but they will not constitute a  balanced mixture of the six molecules. In order to achieve this, we can just repeat the algorithm six times shifting the starting molecule, and using always the same number of obtained coefficients and the same order in the succession of species (in our case: \ch{H2O}, \ch{CO}, \ch{CO2}, \ch{CH3OH}, \ch{NH3}, \ch{CH4}).

\begin{figure}
\centering
\includegraphics[trim={0 12 0 6}, clip, width=0.8\hsize]{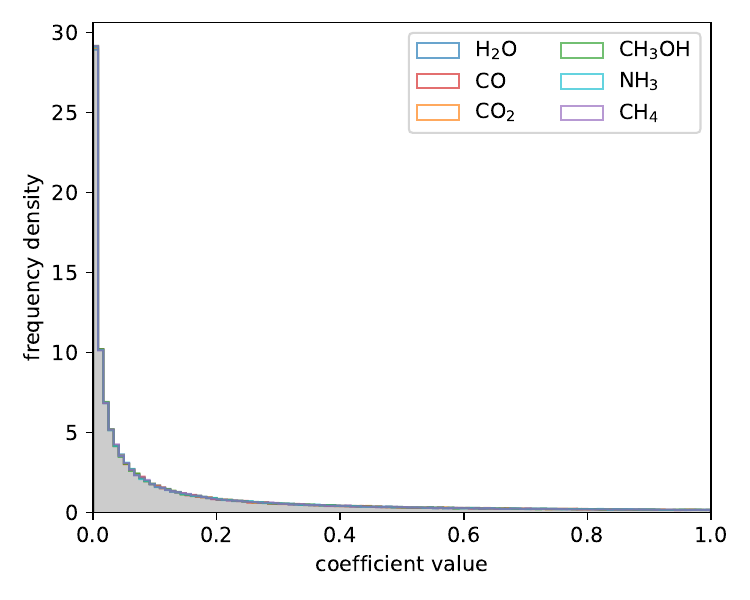}
\caption{Histogram of the generated coefficients using the algorithm mentioned in the text, with 240$\,$000 combinations. In colours we can see the histogram of the population of coefficients for each molecule (they are very similar, and hence difficult to distinguish), while in grey we see the histogram of the values of all the coefficients.}
\label{figure:aice-coeffs}
\end{figure}

This is how we created our synthetic dataset of 282 spectra (creating 47 spectra for each of the 6 repetitions of the algorithm), starting with a random seed equal to 1. If we create a larger number of spectra, we can assess that the obtained sets of coefficients $\{\boldsymbol{c}\}$ are balanced for all the target species, by plotting a histogram of the populations of each coefficient $c_m$. Fig. \ref{figure:aice-coeffs} shows the obtained plot for 240\,000 synthetic coefficients. Indeed, the differences between molecules are negligible, and they arise just from the fact that we use a limited sample of coefficients.

\section{Evolution of the training of AICE}
\label{appendix:aice-training}

\begin{figure}
\centering
\includegraphics[trim={0 12 0 24}, clip, width=0.87\hsize]{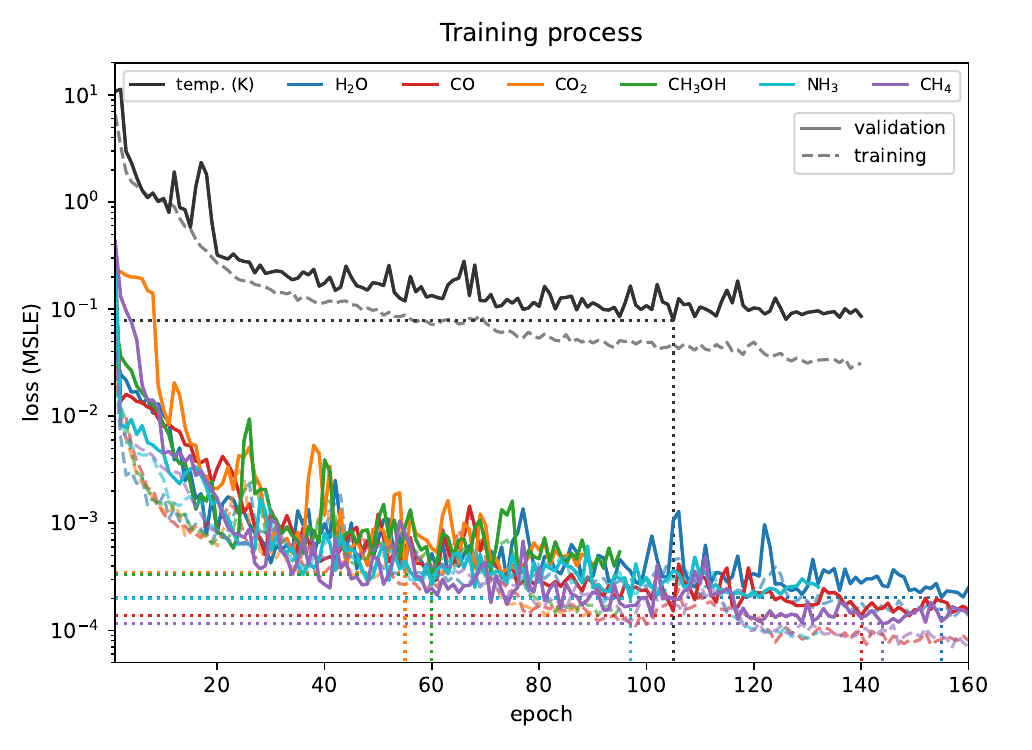}
\caption{Evolution of the training of the set of neural networks  for the first random split.  Dotted lines indicate the minimum values of the loss and the corresponding epochs, for each individual neural network.}
\label{figure:aice-training}
\end{figure}

Figure \ref{figure:aice-training} shows the evolution of the training for the set of neural networks with seed equal to 1. We can see how the losses decrease rapidly at the beginning of the training and slowly stabilise after a maximum amount of 160 epochs. In all cases, the loss in the validation subset is larger than the loss in the training subset, since the optimisation of the model's parameters is done with respect to the training subset. The noisy patterns of the curves are due to the use of mini-batches in the gradient descent. The final weights saved after the training process, for each model, are the ones of the epoch that yielded the minimum loss in the validation subset.  We note that the loss values for the temperature variable are orders of magnitude higher than those of the molecular fraction mainly due to the difference in the range of possible values (from 10 to 100 K for temperature but from 0 to 1 for molecular fraction).

\section{Propagation of uncertainties}
\label{appendix:richvalues}

Throughout the calculations of the paper, there have been several times in which we had to calculate the propagation of uncertainties: for converting the JWST spectra from spectral flux to absorbance, for propagating this last uncertainty in absorbance through the predictions of the neural networks, and also to perform the ratio of the abundances (predicted by AICE or by other works) with respect to \ch{H2O}.

In all these cases, we used the approach of the Python library RichValues,\footnote{\url{https://pypi.org/project/richvalues/}}  similarly as used by \cite{Megias2023}. It consists of using probability density functions (PDFs) to represent the quantities with uncertainties, and then using a Monte Carlo simulation to numerically propagate the uncertainties through any desired function. In the following, we explain the mathematical basis of this approach.

\subsection*{Using PDFs to represent numbers with uncertainties}

Let's consider a variable $x$ with a central value $m$ and an uncertainty $s$, $m \pm s$.  In RichValues, we prefer to base the PDFs in quantiles/percentiles, therefore interpreting $m$ as the median of the PDF (the 50$\,$\% percentile) and ($m - s$, $m + s$) as the $1\,\upsigma$ confidence interval  (68.27 \%) around $m$. One of the reasons for this choice is that the percentiles or quantiles (including the median) are robust to outliers. Another reason is that the median and quantiles transform directly through strictly increasing or decreasing functions; that is, the median of the distribution resulting of applying a function $f$ to an input distribution of values \{$x_i$\} is just the function applied to the median of the original distribution:  ${\rm median}\,f(\{x_i\})$ = $f({\rm median}\,\{x_i\})$ = $f(m_i)$; the same would be valid for any quantile, since strictly increasing or decreasing functions preserve order. Lastly, this interpretation allows for asymmetric uncertainties. This approach is indeed what is used by \cite{Possolo2019} in their work about asymmetric uncertainties.

Then, let's consider a set of $n$ variables $x_j$. To propagate the uncertainties through a function $f$ applied to the input variables, we can draw a sample of a large number of values (e.g. $\gtrsim$$10^4\sqrt{n}$) of each variable $x_j$, apply the function to each of the elements of the samples, and then obtain a central value and an uncertainty for the resulting distribution. This constitutes a type of Monte Carlo simulation, as explained by \cite{Possolo2019}, that allows us to account for asymmetric uncertainties in the input values. To do so, we need two things: an appropriate PDF for converting each variable $x_j$ to a distribution of values, and a proper method to obtain a central value and an uncertainty from the resulting distribution. We will use the median as the central value and the $1\, \upsigma$ confidence interval with respect to it (that includes 68.27 \% of the distribution) to obtain the lower and upper uncertainties. As for the PDF, if the domain of the variable $x$ is ($-\infty$, $\infty$), a proper function is the well-known Gaussian:
\begin{equation}
    \indent \indent
    \label{equation:normal-pdf}
    \mathrm{PDF}(x) \;=\; \frac{1}{\sqrt{\uptau} \; s} \exp \left (- \frac{1}{2} \left (\frac{x-m}{s} \right )^2 \right ) \;,
\end{equation}
with $\uptau \equiv 2\uppi$. In this case, $m$ is at the same time the mean, the median and the mode of the distribution; as for $s$, it is the standard deviation while it also defines the $1\,\upsigma$ (68.27 \%) confidence and credibility intervals. However, if the domain of the variable is not ($-\infty$, $\infty$), this function would be incorrect; for example, if the domain is [0, $\infty$), like the case of molecular abundances, a normal distribution would yield a non-zero probability of negative values, with greater effect the larger the relative uncertainty is.\footnote{Actually, the usual analytic formulas for uncertainty propagation are no longer correct in this case, although they are a good approximation if the relative uncertainty is low. In RichValues, these formulas are used when the uncertainties are low enough.} Therefore, we should use another function as the PDF.

Let's suppose a domain ($b_1$, $b_2$), which defines the possible range of values of our variable. We define the left and right amplitudes, $a_1$ and $a_2$, as the distances between the limits of the domain and the median, that is, $a_1 = m - b_1$, $a_2 = b_2 - m$. Now, as these amplitudes can be different, we will split our desired PDF in two halves, one for $x \leq m$ and other for $x > m$. Then, we will use an amplitude $a$ as a reference, which must be greater than the uncertainty, $a > s$. If the amplitude is quite greater than the uncertainty, $a \gg s$, a good PDF would be just the normal distribution truncated to the domain ($b_1$, $b_2$). However, for amplitudes closer to the uncertainty, it would be considerably incorrect, as the truncation shifts the median of the distribution and modifies the confidence intervals, and thus the uncertainties (it also shifts the mean and modifies the standard deviation). To fix this, we make the following variable change:
\begin{equation}
    \indent \indent
    \label{equation:variable-change}
    \frac{x - m}{a} \; \rightarrow \; \frac{\tilde x - m}{a} \; \equiv \; {\rm arctanh} \left ( \frac{x - m}{a} \right ) \,.
\end{equation}
Using this new variable $\tilde x$ with a normal distribution, we are able to compress the original domain of ($-\infty$, $\infty$) to ($-a$, $a$). Making the corresponding variable changes (see the user's guide,\footnote{\scriptsize{\url{https://github.com/andresmegias/richvalues/blob/main/userguide.pdf}}} section 8.1.2, for more details), we obtain the following formula for the PDF:
\vspace{-0.45cm}
\begin{equation}
    \indent
    \label{equation:normal-alt-pdf}
    \mathrm{PDF}(x) \;=\; \frac{1}{\sqrt{\uptau} \; a \, {\rm arctanh}(s / a)} \frac {\exp \left (- \frac{1}{2} \left (\frac{{\rm arctanh} \left (\frac{x - m}{a} \right )}{{\rm arctanh}(s / a)} \right )^2 \right )}{ 1 - \left ( \frac{x - m}{a} \right )^2} \;.
\end{equation}
We call the corresponding distribution a {\it bounded normal distribution}, and it meets our requirements: the median is equal to $m$, the uncertainty associated with the $1\, \upsigma$ confidence interval (68.27 \%) is $s$, and its range of values is ($m - a$, $m + a$). Since this distribution is symmetric, $m$ is also the mean, although $s$ is no longer the standard deviation.

\begin{figure}
\begin{center}
\includegraphics[trim={0 13 0 0}, clip, width=\hsize]{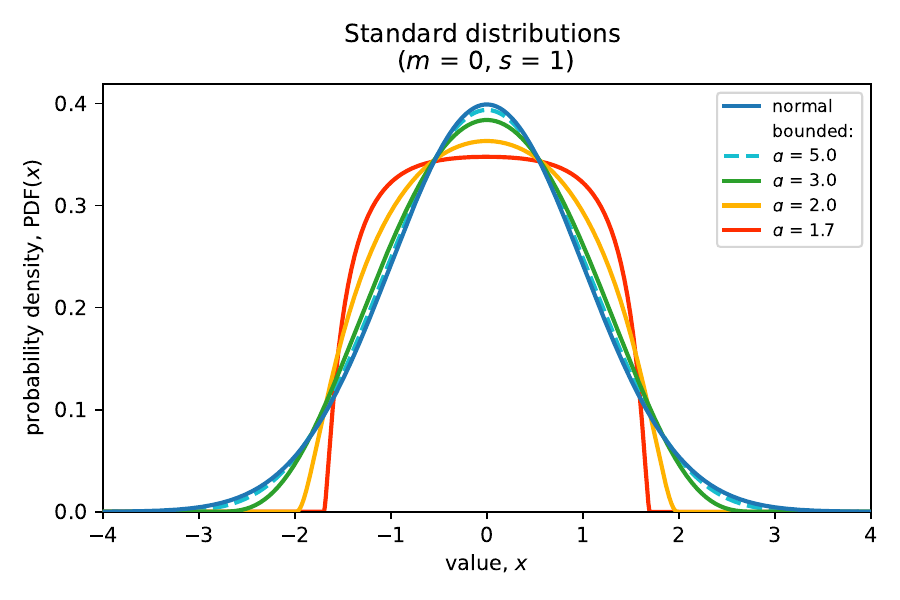}
\end{center}
\caption{Probability density functions for a standard normal distribution and several standard bounded normal distributions with different domains.}
\label{figure:bounded-normal}
\end{figure}

As we can see in Fig. \ref{figure:bounded-normal}, it resembles a normal distribution but with a restricted domain. In fact, for small uncertainties ($s \ll a $) this PDF quickly tends to a Gaussian. On the other hand, for large uncertainties ($s \sim a/2$), the shape of the PDF starts to resemble that of a uniform distribution. Larger normalised uncertainties should not be common, since it would mean a very uninformative value, although the formula could still be used (see the user's guide$\,^{24}$ for more details).

We have defined our PDF for the case of a variable $x$ with a central value $m$, an uncertainty $s$, and a domain of ($m - a_1$, $m + a_2$), building the final PDF with two halves with amplitudes $a_1$ and $a_2$, respectively. In case we had lower and upper uncertainties, $s_1$ and $s_2$, we should just replace $s$ by $s_1$ for the left half of the PDF and by $s_2$ for the right half.

\subsection*{Interpolation for asymmetric distributions}

\begin{figure}
\begin{center}
\includegraphics[trim={0 13 0 0}, clip, width=\hsize]{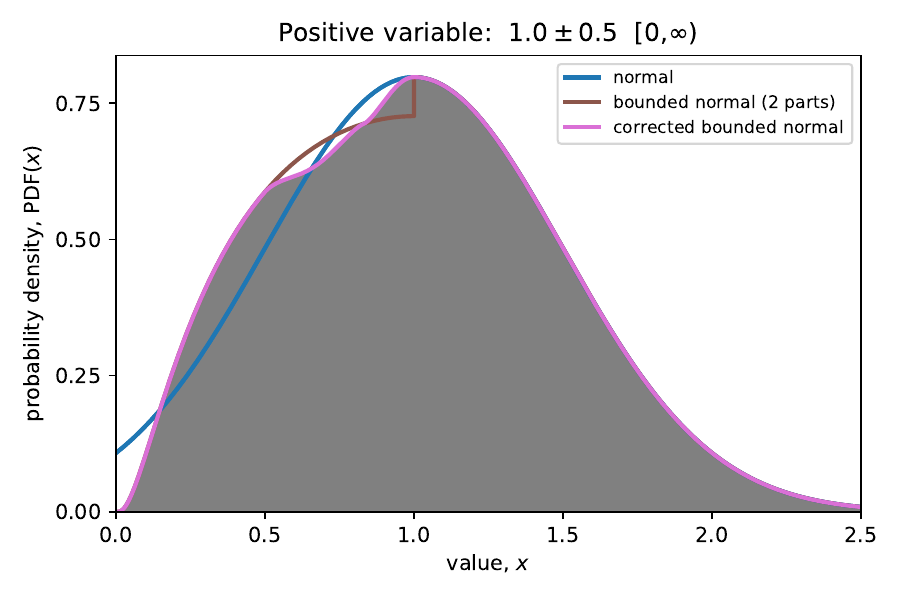}
\end{center}
\caption{Different probability density functions representing a positive variable with a value of $1.0 \pm 0.5$.}
\label{figure:richvalues-example}
\end{figure}

In general, we can have one half of the PDF for $x < m$ with amplitude, $a_1$, and uncertainty, $s_1$, and another half of the PDF for $x \geq m$ with amplitude, $a_2$, and uncertainty, $s_2$. In most cases, there will be a gap in the union of the two functions, as the value at $x = m$ will be different for each half. To avoid this discontinuity, which seems unnatural, we can add a correction to the PDF next to $m$ based on the cosine function. We will not explain it here (for more details, check the user's guide$\,^{24}$), but we can see an example of it in Fig. \ref{figure:richvalues-example}. We call this versatile PDF an asymmetric bounded Gaussian. We note that in this asymmetric case, $m$ is no longer the mean, and neither $s_1$ nor $s_2$ are equal to the standard deviation.

\subsection*{Alternative PDFs for asymmetric uncertainties and others}

In some cases when the amplitudes ($a_1$, $a_2$) are large enough and there is a low or medium asymmetry in the uncertainties ($s_1$, $s_2$), we can use alternative functions to model the PDF of the rich value: the PDFs of a split normal distribution, a log-normal distribution and a generative extreme value (GEV) distribution. We used these three PDFs to represent values with asymmetric uncertainties using the same reasoning as in the work by \cite{Possolo2019}.

In these cases with asymmetries in the uncertainty, the asymmetric bounded Gaussian can produce a bimodal curve, with a second peak due to the cosine interpolation near the median to join the two halves of the bounded Gaussians. If this is the case, one can use these alternative PDFs to avoid that slight bimodality (although it is a PDF that perfectly meets our requirements with respect to to $m$, $s_1$, $s_2$, $a_1$, $a_2$). However, the use of these alternative PDFs comes with an increase of computational cost, since the parameters of these PDFs are not directly $m$, $s_1$, $s_2$, $a_1$, and $a_2$; therefore, we have  to carry out an optimisation to find the proper PDF parameters for each case. This is why this option is disabled by default in RichValues. In any case, the user's guide$\,^{25}$ can be consulted to learn more about these alternative functions and how to use them.

Lastly, our approach with PDFs can also be used to address the case of a variable with an upper/lower limit or even a finite interval, although for the calculations of this paper it is not needed. If it were the case, uniform distributions would be used (either in linear or logarithmic scale), with finite boundaries for infinity ($\pm10^{90}$) and for zero ($\pm10^{-90}$).

\section{Effect of band saturation}
\label{appendix:saturation}

Band saturation can occur when there is a very high column density of material or when the sensitivity of the instrument does not allow us to differentiate the non-zero flux in an absorption feature. In order to explicitly take into account this effect in the training of our neural networks, we performed a test, applying data augmentation simulating the effect of saturation.

\begin{figure}
\begin{center}
\includegraphics[trim={0 13 0 24}, clip, width=\hsize]{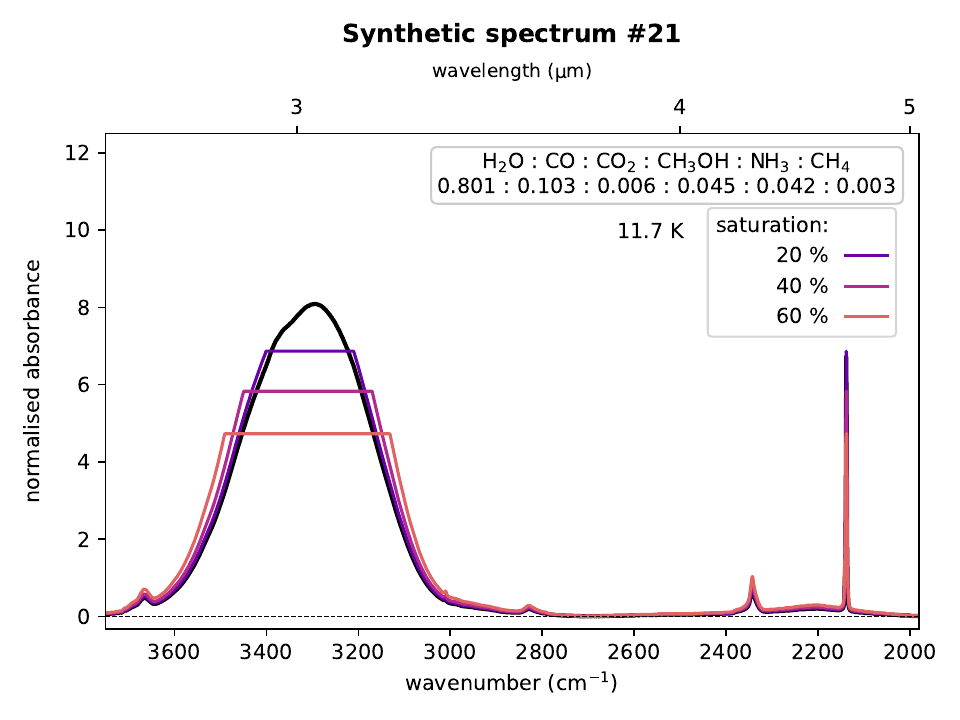}
\end{center}
\caption{Synthetic spectrum \#21 with different levels of simulated saturation: zero (in black) and 20--60 \% (in colours).}
\label{figure:saturation-simulation}
\end{figure}

For every spectrum in our training subset, we have made three copies of it, with different levels of saturation: 20\,\%, 40\,\%, and 60\,\%. To simulate saturation, we crop the curve at values greater than that fraction of the maximum absorbance value, and then we normalise with the new mean absorbance value. An example can be see in Fig. \ref{figure:saturation-simulation}. We note that after the new normalisation after performing the cropping, the whole spectrum gets higher values of absorbance. This could partly explain why the standard version of AICE already provides good estimations of the \ch{H2O}, CO, and \ch{CO2} fractions --especially \ch{CO2}-- for J110621 (see Section \ref{section:band-saturation}).  Another explanation would be that the standard AICE version already focuses on the wings of the bands rather than on the peak. 

\begin{figure}
\begin{center}
\includegraphics[trim={36 15 360 40}, clip, width=0.8\hsize]{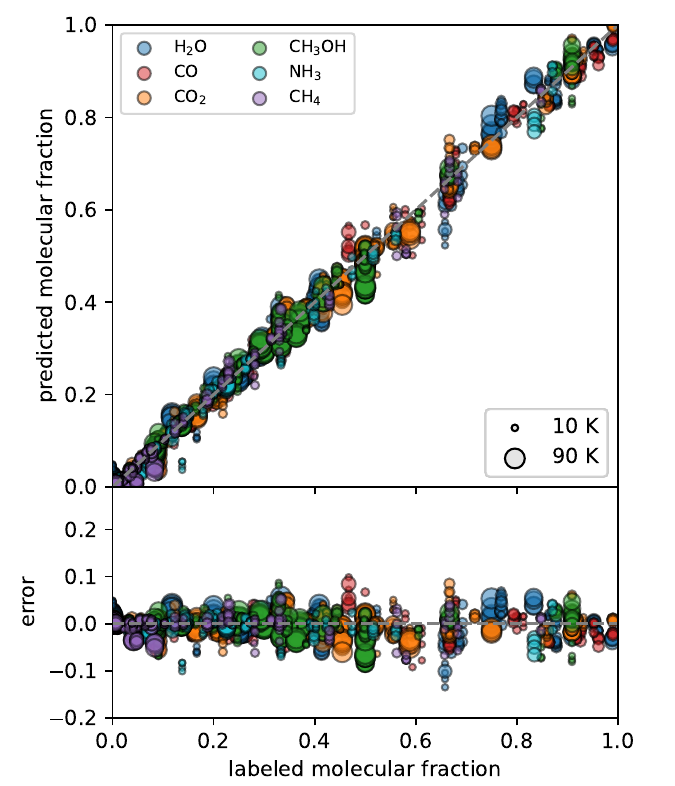}
\end{center}
\caption{Result of the validation for our neural networks after the training with simulated saturation, for the first random split of the dataset. The copies of the spectra with simulated saturation are included here, this is why there are many clusters of points with the same colour and value of labeled molecular fraction. This is equivalent to Fig. \ref{figure:aice-validation}.}
\label{figure:validation-s}
\end{figure}

The  process explained here is applied only to the training subset, not in the validation subset, for each of the different splittings for the different random seeds. We enlarge the batch size within a factor of 4 to account for the increase in the training subset. This makes the training process only around a 30\,\% larger in time, instead of a higher increase, taking in total 33 min to train all the models. After that, we check the performance of our models in the validation (Fig. \ref{figure:validation-s}), obtaining a good correlation between the predicted and actual values, for the molecular fraction and the temperature.

In Section \ref{section:band-saturation} we test this new version with astronomical observations, comparing it with the normal version of our tool. In short, the results by both models are compatible within the uncertainties, although we see some systematic differences. In any case, it does not seem that the standard AICE models are underpredicting any molecular fraction due to band saturation.

\section{Retraining with a smaller spectral range}
\label{appendix:aice-lite}

\begin{figure}
\begin{center}
\includegraphics[trim={36 15 360 40}, clip, width=0.9\hsize]{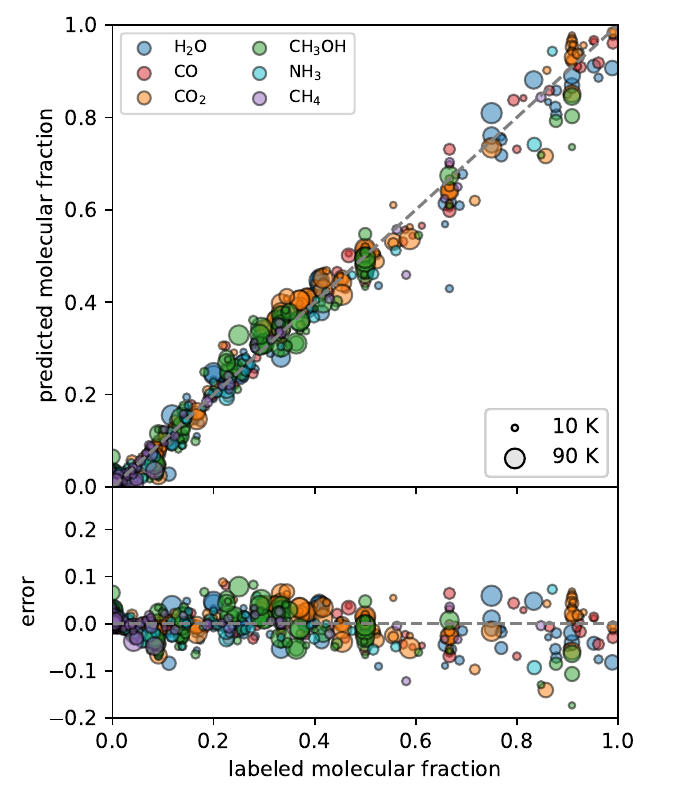}
\end{center}
\caption{Result of the validation for our neural networks after the training in a smaller spectral range (4000--2000 cm$^{-1}$) for the first random split of the dataset. This is equivalent to Fig. \ref{figure:aice-validation}.}
\label{figure:validation-lite}
\end{figure}

As we commented in Section \ref{section:aice-lite}, we retrained our neural networks with a smaller spectral range of 4000--2000 cm$^{-1}$ (2.5--5.0 $\upmu$m), reducing the number of neurons in the first hidden layer from 120 to 90. Despite the reduction in the number of spectral points, we obtained a good correlation between the predicted and actual values in the validation, as shown in Fig. \ref{figure:validation-lite}, although with a higher error. The training of the models took barely the same time as the standard version, around 25 min. In Section \ref{section:aice-lite} we test this new version with astronomical observations, comparing it with the normal version of our tool. The predictions of both versions are compatible within the uncertainties except for \ch{CH3OH} and \ch{NH3}, which are now overestimated.

\section{Automatic band integrator}
\label{appendix:aice-integrator}

\begin{figure*}
\centering
\includegraphics[trim={0 10 0 24}, clip, width=1.0\hsize]{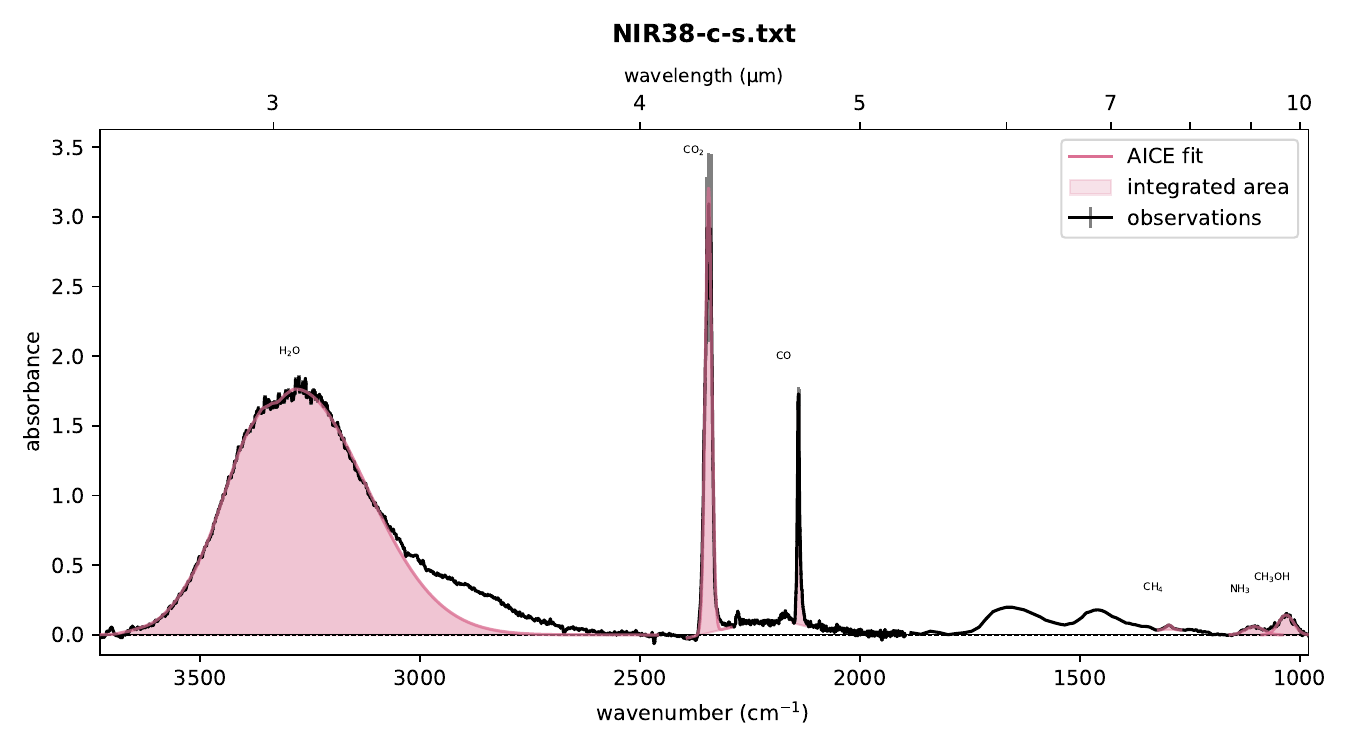}
\caption{Spectra taken by JWST towards NIR38 in the molecular cloud Cha I, converted to absorbance scale after fitting the stellar continuum and the silicate contribution. In pale red are the fitted Gaussians by the band integrator included in AICE (in the case of \ch{H2O}, only the right half of the fit is Gaussian, the left half is a spline; and for CO we directly integrated the observed band, since the shape is not too Gaussian).}
\label{figure:aice-integrator}
\end{figure*}

\renewcommand*{\arraystretch}{1.5}
\begin{table}
\caption{Column densities obtained by the automatic band integrator included within AICE.}
\label{table:aice-integrator}
\centering
\begin{tabular}{lcc}
\hline 
\multirow{2}{*}{\textbf{molecule}} & \multicolumn{2}{c}{\textbf{column density }($10^{18}\,\mathrm{cm}^{-2}$)}\tabularnewline
\cline{2-3} \cline{3-3} 
 & \textbf{NIR38} & \textbf{J110621}\tabularnewline
\hline 
{\bf $\mathbf{H_{2}O}$} & $7.0878 \pm 0.0011$ & $9.0 \pm 0.3$ \tabularnewline 
{\bf CO} & $2.007 \pm 0.008$ & $2.86 \pm 0.10$ \tabularnewline 
{\bf $\mathbf{CO_2}$} & $1.238 \pm 0.020$ & $1.6_{-0.2}^{+0.5}$ \tabularnewline 
{\bf $\mathbf{CH_{3}OH}$} & $0.81_{-0.05}^{+0.07}$ & $1.2_{-0.3}^{+0.8}$ \tabularnewline 
{\bf $\mathbf{NH_{3}}$} & $0.34_{-0.03}^{+0.04}$ & $0.31_{-0.05}^{+0.08}$ \tabularnewline 
{\bf $\mathbf{CH_{4}}$} & $0.171_{-0.007}^{+0.008}$ & $0.211_{-0.011}^{+0.011}$ \tabularnewline 
\hline 
\end{tabular}
\begin{justify}
\footnotesize{{\bf Note.} The band strengths used to scale the integrated area are the ones shown in Table \ref{table:pure-ices}, reported by \cite{Bouilloud2015}.}
\end{justify}
\end{table}
\renewcommand*{\arraystretch}{1.0}

The tool AICE includes a module that allows us to make a simple fit based on Gaussians to individual molecular bands, after writing an initial estimation of the fit parameters in a configuration file. It also allows us to exclude the central part of the curve in order to take into account possible saturation effects on the spectrum to be integrated. We used the band strengths reported in Table \ref{table:pure-ices}. Figure \ref{figure:aice-integrator} shows the result of these fits for the spectrum observed towards NIR38 by \cite{McClure2023}. Table \ref{table:aice-integrator} reports the integrated column densities obtained with this method. The observational uncertainties were propagated to the derived column densities using the Python library RichValues (see Appendix \ref{appendix:richvalues}).

\end{appendix}

\end{document}